\documentclass[12pt]{article}
\usepackage{latexsym} \usepackage{epsf}
\usepackage{epsfig}
\usepackage{a4}
\usepackage{amsfonts}
\usepackage{showlabels}
\usepackage[colorlinks]{hyperref}
\usepackage{color}
\newcommand{\ba}{\begin{array}}
\newcommand{\ea}{\end{array}}
\newcommand{\be}{\begin{equation}}
\newcommand{\ee}{\end{equation}}
\newcommand{\nn}{\nonumber}
\newcommand{\bea}{\begin{eqnarray}}
\newcommand{\ena}{\end{eqnarray}}
\newcommand{\beas}{\begin{eqnarray*}}
\newcommand{\enas}{\end{eqnarray*}}

\renewcommand{\theequation}{\thesection.\arabic{equation}}

\begin{document}

\begin{center}

 {\Large{ Series of the solutions to Yang-Baxter equations: Hecke type matrices and descendant R-, L-operators
 }}

\end{center}

\begin{center}{\bf
Shahane A. Khachatryan \footnote{e-mail:{\sl shah@mail.yerphi.am}
 }}
 \\ A.~I.~Alikhanyan National Science Laboratory (YerPhI), %Yerevan Physics Institute after A.~I.~Alikhanian,
 \\ Alikhanian Br. str. 2, Yerevan 36, Armenia
\end{center}

\begin{abstract}

We have constructed series of the
spectral parameter dependent
 solutions to the Yang-Baxter equations
  defined on the tensor product of
reducible representations with a
 symmetry of quantum (super)algebra.
  These series are produced as descendant
solutions from the $sl_{{q}}(2)$-invariant
Hecke type $R^{r\;r}(u)$-matrices.
 The analogues of the  matrices of Hecke type with the symmetry of  the quantum super-algebra $osp_q(1|2)$ are obtained precisely. For the homogeneous solutions $R^{r^2-1\;r^2-1}$ there are constructed Hamiltonian operators of the corresponding
one-dimensional quantum integrable models, which
 describe rather intricate
 interactions between different kind of spin states. Centralizer operators  defined on the products of the composite states are discussed.  The inhomogeneous
 series of the operators $R^{r\mathcal{R}}(u)$, extended Lax
  operators of Hecke type, also are suggested.

\end{abstract}

{\small\tableofcontents}
\newpage

\section{Introduction}

The Yang-Baxter equations (YBE),  appeared in the early investigations on the problem
of the exactly solvability in  2d statistical
physics, as well as in  $(1+1)$-dimensional scattering theory, and being one of the key relations of the QISM (Quantum Inverse Scattering Method),
 still remain an actual and attractive subject of the statistical and mathematical physics, with expanding area of
   applications \cite{Y}-\cite{LKL}.
 In studying the solutions to the YBE with the symmetry of
 quantum algebra
as  basic constructions the universal $R$ matrices are considered, %Lax operator,
 which can be achieved either by the quantum double principle of Drinfeld or by
the Jimbo's compositions involved algebra invariant matrices - projection operators,
 and different schemes of affinization (or baxterization) are developed for obtaining the spectral
parameter dependent solutions \cite{QGR,VGD,j,j1,grs}.
 It turns out that the range of the spectral
parameter dependent solutions to Yang-Baxter equations with the
given quantum algebra symmetry is richer, than that which can be
constructed via the universal $R$ matrices of the corresponding
quantum algebras. For finding the full class of the symmetric YBE
solutions it is sufficient to consider the $R$-matrix in the
expansion of the whole basis of invariant operators (projectors), which must be specified for the given set of the
representations \cite{j,j1,ji,jim}. In particular, investigating solutions defined on the cyclic and indecomposable representations of the quantum algebra $sl_q(2)$ at roots of unity \cite{SHKH,SHhk},
 we find new solutions and yet a rich variety of the solutions, which are characterised by different structures of decompositions into the projectors and
 as well by additional (spectral) parameters \cite{SHY}.  As a yet another confirmation of the mentioned  observation could be served
the existence of a series of solutions to the YBE  with symmetry of the quantum
super-algebra $osp_q(1|2)$ defined on the spin-irreps, which
differs from the known solutions \cite{kr1,s,SHKKH}, and the discussion done in the Section 2 of this work demonstrates the exact derivation of this series. The similar solutions
(Hecke type $R$-operators), as it is known, exist for the  quantum algebra $sl_q(2)$
\cite{krs,frt,kr,isa,jim}, and this reflects the circumstance that there is
an explicit correspondence between the representations of the
 quantum algebras $sl_q(2)$ and $osp_q(1|2)$,
providing that $q \to i\sqrt{q}$ \cite{k,z,SHKKH,shk}. Then in Section 3 a descendant series of
the mentioned solutions is constructed.
 The non usual
behavior of these $R$-matrices is the reducible character of
 the vector spaces on which the operators act  for
general values of deformation parameter $q$. The integrable models
corresponding to these $R$-matrices describe interactions between
 different spins (Section 4), however the Hamiltonian operator derived
   in accordance to the principles of the Algebraic Bethe Ansatz \cite{FTSZ1,Baxt,KBI} has not the conventional
 form of the superposition of "spin-spin" operators. New formal operators can be proposed for describing
 these nearest neighborhood rather entangled interactions. In  Section 5 an approach determining the centralizer operators defined on the tensor products of the reducible states is developed, necessary and sufficient relations for them are deduced.    By fusion procedure we can
 find out from the Hecke solutions the descendant inhomogeneous $R$-matrices also,  and it turns out that the matrices $R^{r{\mathcal{R}}}$ defined on the tensor product $V^r\otimes U^{\mathcal{R}}$, with $V^r$ being an irrep and $U^{\mathcal{R}}$ being a series of the  reducible states, formed by the truncation of the tensor products of the irreps $V^r$,  %with definite dimensions %${\mathcal{R}}_n$,
 may constitute "Hecke type" matrices by their structure.
  In  Section 6 we sketch the scheme of the obtainment of such $R^{r\mathcal{R}_n}(u)$ operators, defined for each $r$-dimensional irrep and  corresponding series of composite representations with definite  $\mathcal{R}_n$ dimensions. These matrices we can refer to as the series of the "extended Lax operators", as for the case of the fundamental representation of $sl_q(2)$, when $r=2$, they just coincide with the matrix representations of the ordinary Lax operator.     In
  Section 7 the summary and some propositions are presented regarded
 the "extended" R-, L- operators,
and also there are discussed further developments and possible
applications of the integrable structures defined on the composite
representations. In the next part of the Introduction (Section 1),
as well as in the Appendix some preliminary
definitions, descriptions and formulas are presented.  Also in the Introduction some  questions on the baxterization are analyzed.

\paragraph{Quantum super-algebra $osp_q(1|2)$.}
 This graded quantum algebra
 is constituted by the generators $e$, $f$ (odd generators)
and $h$ (even generator), which satisfy to the following
commutation ($[,]$) and anti-commutation ($\{,\}$)  relations
 \bea
 \{e,f\}\equiv e f+f e=[h]_q,\quad  [h,e]=e\quad [h,f]=-f.\label{algebra}
 \ena
Here, as usual, $[a]_q=\frac{q^a-q^{-a}}{q-q^{-1}}$. Sometimes different definitions for the anti-commutation relation in (\ref{algebra}) are used, which are equivalent to this one by simple re-scaling of the generators and the deformation parameter \cite{kr1,s,k,z}.  Co-product
is defined by the following relations
\bea \Delta[e]=e\otimes q^{-\frac{h}{2}}+q^{\frac{h}{2}}\otimes
e,\quad \Delta[f]=f\otimes q^{-\frac{h}{2}}+
q^{\frac{h}{2}}\otimes f,\quad \Delta[h]=h\otimes I+I\otimes
h.\label{coprod}\ena
Here $\otimes$ denotes the graded tensor product, and $I$ is a unity
operator.  The quadratic Casimir operator can be written as
 $c=\left(
(q^{\frac12}+q^{-\frac12})ef-\left[h-\frac12\right]_q\right)^2$. %
Finite-dimensional irreducible representations (irreps) $V^r$,
${\textrm{dim}}[V^r]=r$,  are described by their Casimir
eigenvalues $c_r=[r/2+[(-1)^r+1](\frac{i\pi}{4\log{q}})]_q^2$ and
by "spin" values $j_r=(r-1)/4+[(-1)^r+1](\frac{i\pi}{8\log{q}})$
\cite{k,shk}. Below we shall use the notation
$q_r=[(-1)^r+1]\frac{i\pi}{4\log{q}}$ for the factor arising in
the case of  even dimensional irreps. The odd-dimensional
representations are in the full analogy with the non-deformed
algebra situation, meanwhile the even-dimensional representations
have no well defined limit at $q\to 1$ \cite{k}. The description of the irreducible representations is brought in the Appendix.
%%%%%%%%%%%%%%%%%%%%%%%%%%%%%%%%%%%%%%%%%%%%%%%%%%%%%%%%%%%%%%%%
%%%%%%%%%%%%%%%%%%%%%%%%%%%%%%%%%%%%%%%%%%%%%%%%%%%%%%%%
 %
 The decomposition of the tensor products of two irreps
  is presented by the following linear
combination
\bea V^{r_1}\otimes V^{r_2}=\sum_{r=|r_1-r_2|+1,\; \Delta
r=2}^{r_1+r_2-1}V^r.\label{fusion} \ena
Let us denote
 $j=2j_r-q_r$ and $j_k=2j_{r_k}-q_{r_k}$, $k=1,2$.
The  Clebsh-Gordan
$q$-coefficients (CGC) $C\left(^{j_1 j_2 j}_{i_1\; i_2\;
i}\right)$ are defined by this  decomposition, where it is assumed $\{i_1+i_2=i\}$ and also we
suppose $r_1\leq r_2$,
\bea v^r_{i}=\sum_{i_1=-j_1}^{j_1}C\left(^{j_1 j_2 j}_{i_1\; i_2\;
i}\right)v^{r_1}_{i_1}\otimes v^{r_2}_{i_2}. \label{v12}\ena
Here we have presented  the formulae of  $C\left(^{j_1 j_2 j}_{i_1\; i_2\;
i}\right)$ in such a way to have integer (half-integer) values of
the variables $j,\; i$ for odd (even) dimensional representations
as in the case of $sl_q(2)$-algebra (for details see Appendix A2).
It slightly differs from the notations we have used in \cite{SHKKH},
\cite{shk}. The inverse CG coefficients are defined by
\bea v^{r_1}_{i_1}\otimes v^{r_2}_{i_2}
=\sum_{r=|r_1-r_2|+1}^{r_1+r_2-1}\bar{C}\left(^{j_1\; j_2\;
j}_{i_1\;i_2\;\; i}\right)v^r_{i}. \label{12v}\ena
%
%%%%%%%%%%%%%%%%%%
%%%%%%%%%%%%%%%

%%%%%%%%%%%%%%%%%%%%%%%%%%%%%%%%%%%%%%%%%%%%%%%%%%%%%%%%%%%%%%%%%%%
\begin{figure}[t]
\unitlength=11pt
\begin{picture}(100,10)(5,-1)

\newsavebox{\rmatrix}

\sbox{\rmatrix}{\begin{picture}(10,0)(-2.5,5)
\put(25.5,5){${R}_{ij}%({\scriptsize u_{ij}})
\;=$}\put(30,4){\line(1,1){3}}\put(30,7){\line(1,-1){3}}
\put(29.3,4){\scriptsize$j$}\put(29.3,7){\scriptsize$i$}\put(33.5,4){\scriptsize$i$}
\put(33.5,7){\scriptsize$j$}
\end{picture}}

\multiput(10,4)(0,3){2}{\line(1,0){6}}
\put(7,4){\line(1,1){3}}\put(7,7){\line(1,-1){3}}
\multiput(13,1)(3,0){1}{\line(0,1){9}}
\put(6.5,7){\scriptsize$1$}\put(6.5,3.7){\scriptsize$2$}\put(12.4,0.5){\scriptsize$3$}
\put(16.3,7){\scriptsize$2$}\put(16.3,3.7){\scriptsize$1$}\put(12.4,9.8){\scriptsize$3$}\put(17.2,5){$=$}
\multiput(19.5,4)(0,3){2}{\line(1,0){6}}
\put(19,7){\scriptsize$1$}\put(19,3.7){\scriptsize$2$}\put(22.5,1){\line(0,1){9}}\put(28.5,4){\line(-1,1){3}}
\put(28.5,7){\line(-1,-1){3}}\put(21.9,0.5){\scriptsize$3$}\put(21.9,9.8){\scriptsize$3$}
\put(28.8,7){\scriptsize$2$}\put(28.8,3.7){\scriptsize$1$}
\put(5.5,5){\usebox{\rmatrix}}
\end{picture} \caption{YBE,  ${R}_{ij}$}\label{fig1}
\end{figure}
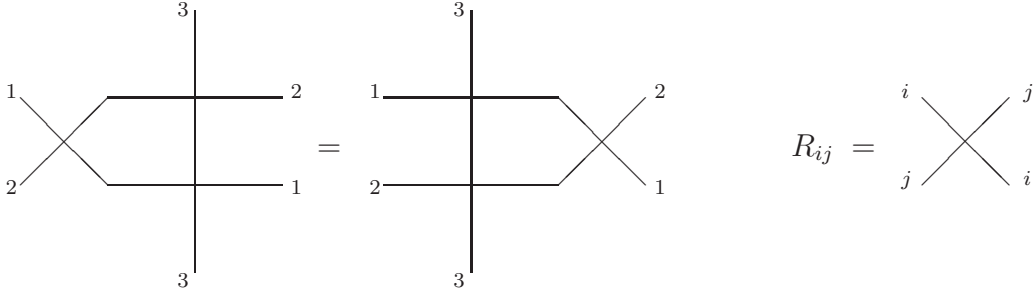
%%%%%%%%%%%%%%%%%%%%%%%%%%%%%%%%%%%%%%%%%%%%%%%%%%%%%%%%%%%%%%%%%%%%%%%%%%%%%%%%%

\paragraph{$R$-matrix and YB equations.}

As a quasi-triangular Hopf algebra this algebra is equipped with
an intertwiner $R$-matrix, which ensures the operation
\be R\Delta[a]=\Delta'[a]R,\quad
\Delta'=\sigma\Delta,\label{rdelta}\ee
where $a$ is an arbitrary element of the algebra and $\sigma$ is
the graded permutation operator acting on the elements of the algebra:
$\sigma\cdot (a\otimes c)=(c\otimes a)$. $R$ satisfies to the
triangle relation (YBE)
\bea R_{12}R_{13}R_{23}=R_{23}R_{13}R_{12}. \label{ybe-const}\ena
Here the right and left sides of the equation are acting on the space
$V_1\otimes V_2\otimes V_3$. $R_{ij}$ is defined  on the product of the
spaces $V_i$ and $V_j$, and acts on the remaining state as
unity operator. In the so-called "check"-formalism it is used
$\check{R}=P R$ - matrix, where $P$ is a graded permutation operator acting on the representation spaces as follows, $P:V_i\otimes V_j=(-1)^{p_i\; p_j}V_j\otimes V_i$. The sign $(-1)^{p_{i}}$ takes into account the grading of the
vector states, the parity $p$ has the values $p_{i}=0/1$ for the even/odd states.
By means of the "check" $R$-matrices the above formulas look like
\bea \check{R}\Delta[a]=\Delta[a]\check{R},\quad
\check{R}_{12}\check{R}_{23}\check{R}_{12}=
\check{R}_{23}\check{R}_{12}\check{R}_{23}.\label{commut}\ena
And note, that in the matrix formulation the "check" YB equations coincide with the non-graded case and do not contain signs, in contrast to the matrix representation of the YBE in (\ref{ybe-const}).
This can be achieved by using the matrix form of the graded tensor products - $(a\otimes c)_{kr}^{ij} = a_k^i c_r^j(-1)^{p_k(p_j+p_r)}$. The solution to YBE as universal  $R$-matrix in terms of the operators $e,\;
f$ and $h$,  is considered by Drinfeld  in
the context of the quantum double principle %(construction)
\cite{VGD}. For the super-algebra $osp_q(1|2)$ it is expressed by
the following formula \cite{s,k}
\bea R^{+}=q^{-h\otimes
h}\sum_{n=0}^{\infty}\frac{(-q^{1/2})^{\frac{1}{2}n(n-1)}(q-q^{-1})^n}{[n]_{+}!}q^{-nh/2}
e^n\otimes f^n q^{nh/2}.\label{r-univ}\ena
Here ${[n]_{+}}=\frac{(-1)^{n-1}q^{n/2}+q^{-n/2}}{q^{1/2}+q^{-1/2}}$. The transpose of this matrix with the change $q\to 1/q$ also is an
intertwiner matrix and is denoted by $R^{-}$.

\paragraph{Spectral parameter dependent $R^{r_1 r_2}(u)$-matrices.}

The appearance of the Yang-Baxter equations in the
quasi-triangular Hopf algebras reflects the connection with the
integrable models \cite{frt}. By means of the affinization of the
quantum groups \cite{QGR,j,grs} it becomes possible to construct
spectral parameter dependent $R(u)$-matrices satisfying to the
equations
\bea
R_{12}(u)R_{13}(u+w)R_{23}(w)=R_{23}(w)R_{13}(u+w)R_{12}(u). \label{ybe}
\ena
As it is well known, the Yang-Baxter equations with spectral
parameter dependence play a crucial role in the theory of the two dimensional
integrable models [1-15]. However the "baxterization" (supplementing with the spectral parameters) of the constant solutions in general is not simple task.

The representation of the matrix $R^{r_1 r_2}(u)$ acting on the
tensor product of two irreps $V^{r_1}\otimes V^{r_2}$ is more
convenient to write in the "check"-formalism, $\check{R}^{r_1
r_2}(u)$. In this case the YB equations are written as
\bea
\check{R}_{12}(u)\check{R}_{23}(u+w)\check{R}_{12}(w)=
\check{R}_{23}(w)\check{R}_{12}(u+w)\check{R}_{23}(u),\label{ybe-check}
\ena
  The commutation relations in (\ref{commut}) and the fusion
rules (\ref{fusion}) give a hint, that the $\check{R}$-matrix must
be a linear superposition of the invariant matrices - projection operators $\check{P}_{r_1
r_2}^{r}$ \cite{j,kr,s},
\bea \check{R}^{r_1
r_2}(u)=\sum_{r=|r_1-r_2|+1}^{r_1+r_2-1}\mathbf{r}_r(u)
{\check{P}}_{r_1 r_2}^{r}, \label{r-sum-p}\ena
 the operator ${\check{P}}_{r_1 r_2}^{r}$ vanishes
  on the spaces $V^{r'}$, $r'\neq r$  in the
decomposition (\ref{fusion}), and acts  on the space
$V^r$, mapping (imaging) it to the space $V^r$ of the decomposition
$V^{r_2}\otimes V^{r_1}$ \cite{SHKKH}. For the homogeneous case the matrices ${\check{P}}_{r_1 r_1}^{r}$  are the
ordinary projection operators $P^{r}$  acting on $V^r$ as
unity
operator, $P^r V^{r'}=\delta_{rr'}V^{r'}$.
  For the quantum super algebra $osp_q(1|2)$ the $\mathbf{r}_r(u)$-functions (which are the polynomials as for the case of $sl_q(2)$ algebra)  when both of
$r_1$ and $r_2$ are odd or even dimensional irreps, can be found e.g. in
\cite{j,kr,s,SHKKH}. The general case of $r_1\neq r_2{(\;\textrm{mod})}\;2$ is
analyzed in detail in \cite{SHKKH}.

All the solutions with higher spin representations ($r_{1,2}>2$) in the series
(\ref{r-sum-p}), which satisfy to (\ref{ybe-check}) can be obtained
by another way, by means of the so called "fusion" technique or "descendant" procedure
\cite{krs,j} from the fundamental $R^{2\;2}(u)$ solution.
This is the consequence of two factors. One factor is that there
is a point $u_0$ for which $\check{R}^{r_1\;r_2}(u_0)$ is proportional to the
projector with the maximal spin $\check{P}^{(r_1+r_2-1)}_{r_1\;r_2}$. The second one is that the expression on left (or right) hand
 side of the YBE (\ref{ybe}) itself can serve as a solution of
YBE  at $u=u_0$ (or at any point $u=\bar{u}_0$, for which $\check{R}(\bar{u}_0)$ is a projector or direct sum of the projectors, and thus has the property $\check{R}(\bar{u}_0)\check{R}(\bar{u}_0)=\check{R}(\bar{u}_0)$ for the homogeneous case; for the general case the permutation must be taken into account in the multiplication of the projectors), in the form of $\check{R}^{(r_1\times r_2)\;r_3}$-matrix acting on the
product of the vector spaces $V^{r_1\times r_2}\otimes V^{r_3}$,
where $V^{r_1\times r_2}=(V^{r_1}\otimes V^{r_2})$.

Note, that as a rule, in this paper we are using the upper indexes
for R-matrices  for denoting the dimensions of the representation
spaces on which the $R$ matrices are acting and the down indexes
for denoting the positions of the states.

\paragraph{Baxterization: some observations.}

For the two-dimensional fundamental representations of
 $sl_q(2)$ (definition of this algebra is brought in A3), the spectral
parameter dependent $R(u)$-matrix (\ref{r-sum-p}) defined on them
 can just be presented by the following sum, up to multiplication by an arbitrary function (see for example \cite{grs,s})
\bea R^{2\;2}(u)=q^u {R^+}^{2\;2}_f -
q^{-u}{R^-}^{2\;2}_f.\label{rnn}\ena
This is true also for the two dimensional representations of
 the $osp_q(1|2)$ super-algebra \cite{SHKKH}.   From this form it means that the matrices $R^{\pm}_f$ on the
fundamental representations besides of the ordinary constant YBE
\bea
R^{\pm}_{f12}R^{\pm}_{f13}R^{\pm}_{f23}&=&R^{\pm}_{f23}R^{\pm}_{f13}R^{\pm}_{f23},\label{ye1}\ena
  must satisfy also the other equations, which are
\bea
%R^{\pm}_{f12}R^{\pm}_{f13}R^{\pm}_{f23}&=&R^{\pm}_{f23}R^{\pm}_{f13}R^{\pm}_{f23},\\
R^{\pm}_{f12}R^{\pm}_{f13}R^{\mp}_{f23}&=&R^{\mp}_{f23}R^{\pm}_{f13}R^{\pm}_{f23},\nn\\
R^{\pm}_{f12}R^{\mp}_{f13}R^{\mp}_{f23}&=&R^{\mp}_{f23}R^{\mp}_{f13}R^{\pm}_{f23},\label{ye2}\\\nn
R^{+}_{f12}R^{-}_{f13}R^{+}_{f23}-R^{-}_{f12}R^{+}_{f13}R^{-}_{f23}&=&-
R^{-}_{f23}R^{+}_{f13}R^{-}_{f23}+R^{+}_{f23}R^{-}_{f13}R^{+}_{f23}.
\ena

For the representations with the higher spins the expansion of
$R^{r_1\;r_2}(u)$ (we suppose $r_1\leq r_2$) to the series in
terms of the parameter $q^{u}$ can contain more terms, i.e. at
$r_1>2$
\bea R^{r_1\;r_2}(u)=q^{u(r_1-1)} {R}^{(r_1)} +q^{u (r_1-3)}
{R}^{(r_1-1)}+\cdots + q^{-u(r_1-1)}{R}^{(1)},\label{r12q}\\
{R}^{(r_1)}\approx{R^+}^{r_1 r_2},\quad {R}^{(1)}\approx{R^-}^{r_1
r_2}.\ena
The last two matrices (${R^+}^{r_1 r_2},\;{R^-}^{r_1 r_2} $) are
the braid limits of the corresponding $R$-matrix  and satisfy to the
equations (\ref{ye1}). The matrices $R^{2\;r}(u)$ constitute the
matrix representations of the Lax operator $L(u)$
\cite{j,kr,SHKKH}, which keeps the form $L(u)=q^u L^--q^{-u}L^-$. The matrices $R^{r_1\;r_2}(u)$ which are obtained by the fusion method from the YBE solutions $R^{2\;2}(u)$ defined on the fundamental representations  admit the form (\ref{r12q}) with non vanishing terms $q^{u(2p-r_1-1)}{R}^{(p)}$ for all variables $p$, meanwhile  the Hecke type homogeneous matrices $R^{r\;r}(u)$ contain only the terms with braid limit matrices $R^{\pm}$ \cite{krs,j,kr,j1,isa,jim}. In this article we shall demonstrate the existence of the series $R^{r\; r}(u)$ of YBE solutions with  minimal number of terms in the  decomposition (\ref{r12q}) also for the quantum super-algebra $osp_q(1|2)$. Then by the fusion method the descendant solutions of such kind matrices will be considered, which for the case $r>2$ will be defined on the reducible representations.

%{\color{red}{$\ast\ast\ast\ast\ast\ast\ast\ast\ast\ast\ast\ast\ast\ast\ast\ast\ast$}
%\color{blue}{$\ast\ast\ast\ast\ast\ast\ast\ast\ast\ast\ast\ast\ast\ast\ast\ast\ast$}
%\color[rgb]{1.00,0.44,0.18}{$\ast\ast\ast\ast\ast\ast\ast\ast\ast\ast\ast\ast\ast\ast\ast\ast\ast$}}

Let us see what relations are imposed by the YBE (\ref{ybe})
on the expansion matrices in (\ref{r12q}). Denoting
$R^{r_i\;r_j}(u)=\sum_{p=1}^{r_i}q^{u(2p-r_i-1)} {R}^{(p)}_{ij}$
($r_i\leq r_j$), and supposing $r_1\leq r_2 \leq r_3$, we have
from the YBE
\bea \sum_{p=1}^{r_1}q^{u(2p-r_1-1)}
{R}^{(p)}_{12}\sum_{k=1}^{r_1}q^{(u+v)(2k-r_1-1)} {R}^{(k)}_{13}
\sum_{t=1}^{r_2}q^{v(2t-r_2-1)} {R}^{(t)}_{23}=\nn\\
\sum_{t=1}^{r_2}q^{v(2t-r_2-1)}
{R}^{(t)}_{23}\sum_{k=1}^{r_1}q^{(u+v)(2k-r_1-1)}
{R}^{(k)}_{13}\sum_{p=1}^{r_1}q^{u(2p-r_1-1)} {R}^{(p)}_{12}, \ena
the following set of the equations
\bea \sum_{p=1}^{r_1}{R}^{(p'-p)}_{12}{R}^{(p)}_{13}
{R}^{(p''-p)}_{23}=\sum_{p=1}^{r_1}{R}^{(p''-p)}_{23}{R}^{(p)}_{13}{R}^{(p'-p)}_{12},\label{r123q}\\
\nn 1\leq (p'-p)\leq r_1, \;\;\; 1\leq (p''-p)\leq r_2,\;\;\;\;\; {\mathrm{i.e.}}\;\;\;\;\;
p'\in [2,2r_1],\;\;\; p''\in [2,r_1+r_2].\ena
 These equations can be investigated step by step, starting from the braid limit matrices. Of course the expansion (\ref{r12q}) is valid for the trigonometric (or rational, at $q\to 1$) solutions. For more general solutions infinite series must be considered, i.e. the summation must not be limited by $r_1$ in (\ref{r12q}), it must be extended to $\infty$.

  The solutions to the equations (\ref{r123q}) are not unique. As an example we
can consider the R-matrices with the symmetry of the quantum
algebra $sl_q(2)$ for the case $r_{1,2}=3$.
 There are three spectral-parameter dependent solutions
 $R^{33}(u)$ to the homogeneous YBE. Presenting them in the form
 $\check{R}^{33}(u)=q^u\check{R}^++\check{R}^0+q^{-u}\check{R}^{-}$, the first solution
 can be written as
\bea \check{R}^{33}_1(u)=\frac{1}{\bar{a}}\Big(q^u(q^{-3}P^5 -q
P^3+q^3 P^1)+\check{R}^0_1+q^{-u}(q^{3} P^5-q^{-1} P^3+q^{-3}
P^1)\Big),\nn\\
\bar{a}=(q^2-q^{-2})(q-q^{-1}),\quad
\check{R}^0_1=-(q+q^{-1})(P^5+P^1)+(q^3-q^{-3})P^3,\label{r31}\ena
 which is the case $n=3$ of the universal solution
 $\check{R}^{nn}(u)$ \cite{j}. The next solution, associating with the Berman-Wenzl-Murakami
algebra \cite{isa}, reads as
\bea \check{R}^{33}_2(u)=\frac{1}{a'} \Big(-q^u q^{-2}(q^{-3}P^5
-q P^3+q^3 P^1)+\check{R}^0_2+q^{-u}q^2(q^{3} P^5-q^{-1}
P^3+q^{-3} P^1)\Big),\nn \\a'=(q^2-q^{-2})(q^3+q^{-3}),\quad
\check{R}^0_2=(q^5-q^{-5})(P^3+P^1)+(q^{-1}-q)P^5. \label{r32}\ena
This matrix has the same braid limits $R^{\pm}$ as the previous
one. We see that having
 the same braid limit constant matrices $R^{\pm}$ it is possible to
 construct different spectral parameter dependent solutions (\ref{r31},\ref{r32}). And the third solution
\bea \check{R}^{33}_3(u)=\frac{1}{1+a}\Big(q^u[P^5+ P^3+a
P^1]+q^{-u}[a(P^5+ P^3)+ P^1]\Big),\label{r33}\ena
$a=\frac{\sqrt{q^4+q^{-4}-1+2q^2+2q^{-2}}+q^2+1+q^{-2}}{\sqrt{q^4+q^{-4}-1+2q^2+2q^{-2}}-(q^2+1+q^{-2})}$,
is just the solution belonging to the so called Hecke type
\cite{grs,j} series.
All the solutions are brought in the normalized form
$\check{R}^{33}(0)=I$.

    The
 $osp_q(1|2)$-invariant
$R^{33}$-matrices, which are equivalent to the solutions
(\ref{r31}, \ref{r32}) are discussed in details in
 \cite{SHKKH}.  And it is known,  that in contrast to the case (\ref{r31}), the solution (\ref{r32})
 has no generalization for higher dimensional irreps \cite{SHKKH,jim}.

  The $osp_q(1|2)$-invariant analog of the solution
(\ref{r33}) will be obtained in the next section. %Then in the following sections
 We see, that this solution  has the simplest decomposition  with two constant braid limit $R^{\pm}$-matrices and $\left[u(r-1)\to u\right]$ (\ref{r12q}), i.e. it preserves the form of (\ref{rnn}). Note that the spectral parameter dependent YBE and R-matrices are defined up to the re-normalization of the spectral parameter.

\section{A series of  Hecke type homogeneous solutions to YBE  with  $osp_q(1|2)$-symmetry}
 \addtocounter{section}{0}\setcounter{equation}{0}

 One can try to make the generalization of the fundamental representation (\ref{rnn})
for the higher dimensional cases  in such a way,
that to keep the form  $R(u)=R^{(+)} q^u-R^{(-)} q^{-u}$. %(as it is for (\ref{rnn})).
 It is known, that the Hecke type R-matrices,
i.e. the matrices, which satisfy the Hecke relation
$(\check{R}-q)(\check{R}^{-1}+q)=0$, after "baxterization" obtain
the mentioned form (e.g. \cite{isa} and citations therein). And surely
for higher dimensions these  ${R}^{(+/-)}$-matrices do not
coincide with the braid limit matrices ${R}^{+/-}$ obtained from
the universal R-matrix.

Although for the quantum super-algebras $osp_q(1|2N)$ the role of
the  Hecke algebra is played by the Birman-Wenzl-Murakami algebra, taking into account the equivalence of the quantum algebras $sl_q(2N)$ and the quantum super-algebras $osp_q(1|2N)$
 in respect to their representation spaces \cite{z},  one can also
expect the existence of the series of Hecke type R-matrices with the symmetries of these
super-algebras.
 Now let us concentrate our attention on the case of $osp_q(1|2)$.
For the irreps with the
dimensions $r=2$ and $r=3$ all the solutions $\check{R}^{rr}$ one
can obtain by direct matrix calculations and verify that there are the
counterparts of the solutions (\ref{rnn}, \ref{r33}).
In the case of the general $r$ let us look for the solution of
(\ref{ybe-check}) in a special form
\bea \check{R}^{r r}(u)=\sum_{r'=3}^{2r-1} {{P}}_{r
r}^{r'}+\bar{f}(u)P^1= I^r\otimes I^r+f(u)P^1 \label{r-f-p},\ena
as the solutions at $r=2,\;3$
admit  such expansion (and this is valid in the case of $sl_q(2)$ invariant Hecke type operators, too).
 Here $I^r$ is  unity operator in the space $V^r$ and
$f(u)=\bar{f}(u)-1$, as $\sum_{r'=1,\;\triangle r'=2}^{2r-1} {P}^{r'}=I^r\otimes I^r$.

Note, that dealing with the homogeneous matrices  for simplicity  we  use the notations $P^{r}_{r_i r_k}\equiv P^{r}_{ik}$ ($r_i=r_k$), and in some cases just $P^r$, without the  indexes $i,\; k$  denoting the spaces.

It is possible to derive the functions $f(u)$ by various methods exploiting the algebra
relations. Here we shall demonstrate an explicit
computation in a rather detailed way, using the Clebsh-Gordan
coefficients.

The procedure is standard.
  The right and left sides of the YBE are acting on the space $V^r\otimes V^r\otimes V^r$.
Let us  take an arbitrary vector state in that space, suppose
$v^r_{k}\otimes v^r_p \otimes v^r_t$. The projector $P_{i j}^1$
acts as non-vanishing (unity) operator only on this kind of
the products - $v^r_{k}\otimes v^r_{-k}$, which %
 have $0$-value of the operator $h$. Using definitions in
(\ref{v12}, \ref{12v}) and denoting $j_0=(r-1)/2$, we can write
\bea \nn (P^1)_{12}\cdot v^r_{k}\otimes v^r_p \otimes
v^r_t=(P^1\otimes I)\cdot v^r_{k}\otimes v^r_p \otimes
v^r_t=\\\delta_{k,-p}\bar{C}\left(^{j_0\; j_0\; 0}_{k\;-k\;\;
0}\right)\sum_{i=-j_0}^{j_0}C\left(^{j_0\; j_0\; 0}_{i\;-i\;\;
0}\right) v^r_{i}\otimes v^r_{-i} \otimes v^r_t,\\\nn (P_{r
r}^1)_{23}\cdot v^r_{k}\otimes v^r_p \otimes v^r_t=(I\otimes P_{r
r}^1)\cdot v^r_{k}\otimes v^r_p \otimes
v^r_t=\\\delta_{p,-t}\bar{C}\left(^{j_0\; j_0\; 0}_{p\;-p\;\;
0}\right)\sum_{i=-j_0}^{j_0}C\left(^{j_0\; j_0\; 0}_{i\;-i\;\;
0}\right) v^r_{k}\otimes v^r_{i} \otimes v^r_{-i}.\ena
Taking into account these relations let us write down the
non-trivial equations which follow from the action of the left
and right hand sides of the YBE with the $R$-matrices described by
(\ref{r-f-p}). For definiteness we take $p=-k$
\bea \label{eq-f}\left\{f(u)+f(w)-f(u+w)+f(u)f(w)+\right.\\\left.
f(u)f(w)f(u+w)C\left(^{j_0\; j_0\; 0}_{t\;-t\;\;
0}\right)\bar{C}\left(^{j_0\; j_0\; 0}_{t\;-t\;\;
0}\right)C\left(^{j_0\; j_0\; 0}_{-t\;t\;\;
0}\right)\bar{C}\left(^{j_0\; j_0\; 0}_{-t\;t\;\;
0}\right)\right\}=0.\nn
%C\left(^{j_0\; j_0\; 0}_{k\;-k\;\; 0}\right)
\ena
This equation has solution for all $t$-s, if the factor
$\chi(j_0)\!=\!C\left(^{j_0\; j_0\; 0}_{t\;-t\;\;
0}\right)\!\bar{C}\left(^{j_0\; j_0\; 0}_{t\;-t\;\;
0}\right)\!C\left(^{j_0\; j_0\; 0}_{-t\;t\;\;
0}\right)\!\bar{C}\left(^{j_0\; j_0\; 0}_{-t\;t\;\; 0}\right)$ is
not dependent from the value of $t$. For defining that factor, let us write explicitly the CG-coefficients for the
given case
\bea C\left(^{j_0 j_0\; 0}_{i\; -i\;
0}\right)=\prod_{i'=-j_0+1}^{i}
\frac{-(-1)^{p_{i'}}q^{-\frac{1}{2}-q_{r_0}}\beta^{r}_{i'-1}}{\beta^{r}_{-i'}}
C\left(^{\;\;\;j_0\; j_0 \;0}_{-j_0\; j_0\;
0}\right),\\\label{cj0}
\bar{C}\left(^{j_0\;j_0\;0}_{i\;-i\;0}\right)=
(-1)^{p^{j_0}_{i}p^{j_0}_{-i}}\varepsilon^0_0\varepsilon^{j_0}_{i}\varepsilon^{j_0}_{-i}\prod_{i'=-j_0+1}^{i}\frac{-(-1)^{p_{i'}}
q^{-\frac{1}{2}-q_{r}} \beta_{i'-1}^{r}}{\beta_{-i'}^{r}}
{C}\left(^{j_0\;j_0\;0}_{-j_0\;j_0\;0}\right).
\ena
Hence, using that $(\frac{\beta^{r}_{i-1}}{\beta^{r}_{-i}})^2=1$
and $(\varepsilon^{j}_{i})^2=1$  for all possible $i$, we arrive
at
\bea C\left(^{j_0\; j_0\; 0}_{i\;-i\;\;
0}\right)\bar{C}\left(^{j_0\; j_0\; 0}_{i\;-i\;\;
0}\right)C\left(^{j_0\; j_0\; 0}_{-i\;i\;\;
0}\right)\bar{C}\left(^{j_0\; j_0\; 0}_{-i\;i\;\; 0}\right)=\\\nn
\prod_{i'=-j_0+1}^{i} q^{-1-2q_{r}} \prod_{i'=-j_0+1}^{-i}
q^{-1-2q_{r}}\times%\\\nn
 C\left(^{\;\;\;j_0\;
j_0 \;0}_{-j_0\; j_0\; 0}\right)^4=
q^{-(1+2q_{r})2j_{0}}C\left(^{\;\;\;j_0\; j_0 \;0}_{-j_0\; j_0\;
0}\right)^4\ena
If to choose the $j_0$-state having even norm, then the state
$\{v_0^{1}\}$ would have $0$ parity, and combining the relations
(\ref{ccu}) and (\ref{cj0}) we deduce
\bea C\left(^{\;\;\;j_0\; j_0 \;0}_{-j_0\; j_0\; 0}\right)^2\sum_i
(-1)^{p^{j_0}_{i}p^{j_0}_{-i}}q^{-(1+2q_{r})(i+j_0)}=1.\ena
When $r$ is odd, then
$(-1)^{p^{j_0}_{i}p^{j_0}_{-i}}=(-1)^{j_0+i}$, $q_{r}=0$ and\\
$\sum_{i=-j_0}^{j_0}
(-1)^{p^{j_0}_{i}p^{j_0}_{-i}}q^{-(1+2q_{r})(i+j_0)}=\sum_{i=-j_0}^{j_0}
(-q)^{-(i+j_0)}=(-q)^{-j_{0}}[2j_{0}+1]_{iq^{1/2}}$.\\
 When $r$ is even, then $(-1)^{p^{j_0}_{i}p^{j_0}_{-i}}=1$, and again\\
 $\sum_{i=-j_0}^{j_0}
(-1)^{p^{j_0}_{i}p^{j_0}_{-i}}q^{-(1+2q_{r})(i+j_0)}=\sum_{i=-j_0}^{j_0}
q^{-(1+\frac{i\pi}{\log{q}})(i+j_0)}=(-q)^{-j_{0}}[2j_{0}+1]_{iq^{1/2}}$.

So, we have, using the relation $(-1)^{2j_{0}}=1$ for odd
dimensional representations, and
$q^{-(1+2q_{r})2j_{0}}=(-q)^{2j_{0}}$ for even dimensional ones,
\bea \label{fcr}q^{-(1+2q_{r})2j_{0}}C\left(^{\;\;\;j_0\; j_0
\;0}_{-j_0\; j_0\;
0}\right)^4=q^{-(1+2q_{r})2j_{0}}/\left((-q)^{-j_{0}}[2j_{0}+1]_{iq^{1/2}}\right)^2=
\frac{1}{[2j_{0}+1]_{iq^{1/2}}^2}.\ena
%
%Note that $2j_{0}+1=r_0$.

 The solutions of the equations
(\ref{eq-f}) so have dependence only from the dimension of the
representation $V^r$. Equations can easily be solved by passing to
the corresponding differential equations, expanding the
expressions around a fixed point, e.g. at $w=0$. We can write finally the
spectral parameter dependent Hecke type solutions as
\bea \check{R}^{r r}(u)= I^r\otimes I^r+f(u)P^1,\qquad
f(u)=\frac{2}{-1+\sqrt{1-\frac{4}{[r]_{iq^{1/2}}^2} }\coth{a
u}}.\label{fs}\ena
Here $a$ is an arbitrary number. The "braid"-limits
 $u\to\pm\infty$ of (\ref{fs}) %(besides of the simple case $f(u)=0$)
coincide with the
corresponding Hecke type constant solutions, satisfying to (\ref{ye1}).

The investigation of the case with symmetry of quantum algebra
$sl_q(2)$ %
would differ from this consideration
only by the gradings of the states and the spin values of the
even-dimensional irreps. For the algebra $sl_q(2)$ the Hecke type
solutions have been discussed in the works \cite{baxter,j1,isa}.
The purpose of this section has been to insist that such kind of
series of the solutions exists for the super-algebra $osp_q(1|2)$,
thus proofing that there is a full correspondence between the YBE
solutions with the symmetries of $sl_q(2)$ and $osp_q(1|2)$.

\section{Descendant $R^{(r^2-1)\times(r^2-1)}$-matrices}
 \addtocounter{section}{0}\setcounter{equation}{0}

%%%%%%%%%%%%%%%%%%%%%%%%%%%%%%%%%%%%%%%%%%%%%%%%%%%%%%%%%%%%%%%%%%%
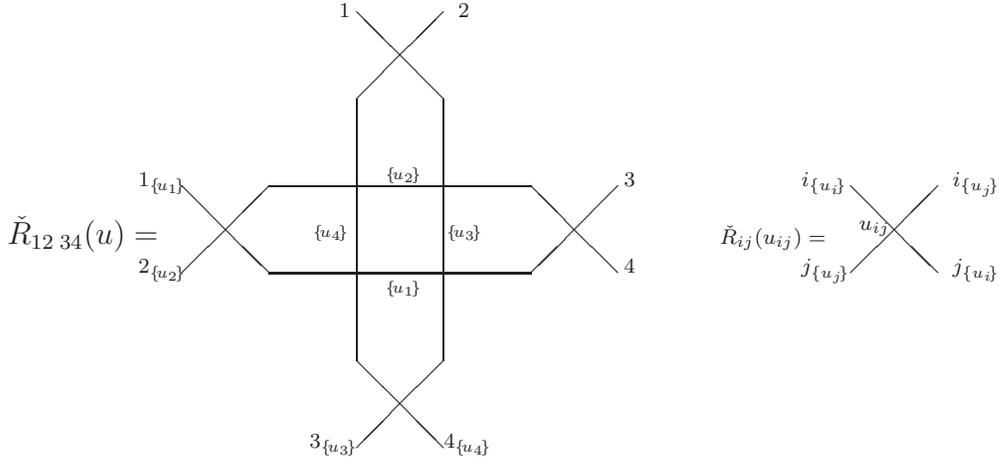
\begin{figure}[t]
\unitlength=11pt
\begin{picture}(100,15)(-1,-1)
\put(1,5){$\check{R}_{12\; 34}(u)=$}
\multiput(10,4)(0,3){2}{\line(1,0){9}}
\put(7,4){\line(1,1){3}}\put(7,7){\line(1,-1){3}}
\multiput(13,1)(3,0){2}{\line(0,1){9}} \put(13,1){\line(1,-1){3}}
\put(16,1){\line(-1,-1){3}}
\put(22,4){\line(-1,1){3}}\put(22,7){\line(-1,-1){3}}
\put(13,10){\line(1,1){3}}\put(16,10){\line(-1,1){3}}
\put(5.5,7){\scriptsize$1_{\{\!u_1\!\}}$}
\put(5.5,4){\scriptsize$2_{\{\!u_2\!\}}$}
\put(11.4,-2){\scriptsize$3_{\{\!u_3\!\}}$}
\put(16,-2){\scriptsize$4_{\{\!u_4\!\}}$}
\put(14,3.5){\scriptsize${}_{\{\!u_1\!\}}$}
\put(14,7.4){\scriptsize${}_{\{\!u_2\!\}}$}
\put(11.5,5.4){\scriptsize${}_{\{\!u_4\!\}}$}
\put(16.1,5.4){\scriptsize${}_{\{\!u_3\!\}}$}
\put(22.2,7){\scriptsize$3$}\put(22.2,4){\scriptsize$4$}\put(12.4,12.8){\scriptsize$1$}\put(16.5,12.8){\scriptsize$2$}
\put(25.5,5){\scriptsize$\check{R}_{ij}({{u_{ij}}})
=$}\put(30,4){\line(1,1){3}}\put(30,7){\line(1,-1){3}}
\put(28.3,4){\scriptsize$j_{\{u_j\!\}}$}\put(28.3,7){\scriptsize$i_{\{u_i\!\}}$}\put(33.5,4){\scriptsize$j_{\{u_i\!\}}$}
\put(33.5,7){\scriptsize$i_{\{u_j\!\}}$}
\put(30.3,5.5){\scriptsize$u_{ij}$}
\end{picture} \caption{ $\check{R}_{12\; 34}$ and
$\check{R}_{ij}$
 }\label{figrr}
\end{figure}
%%%%%%%%%%%%%%%%%%%%%%%%%%%%%%%%%%%%%%%%%%%%%%%%%%%%%%%%%%%%%%%%%%%%%%%%%%%%%%%%%

We can try to construct  descendant
 $R$-matrices corresponding to the
discussed $r\times r$-dimensional solutions. As it is mentioned already, in the standard
fusion (descendant) procedure developed for the algebras under consideration \cite{krs,j} one leans on the
property that there is a point $u_0$ at which the usual solution
$\check{R}^{r_1\;r_2}(u)$ (\ref{r-sum-p}) defined on the tensor product
of the spin-irreps $V^{r_1}\otimes V^{r_2}$ is proportional to the
projector $\check{P}^{r_1+r_2-1}_{r_1\;r_2}$ on the space with the maximal spin in
the fusion
 (\ref{fusion}). Thus from the
matrices $R^{r_1\;r_2}$ and $R^{r_3\;r_4}$ satisfying the
YBE one can construct the matrix
$R^{(r_1+r_2-1)\;(r_3+r_4-1)}$ on the product with  maximal spins (Fig. 2). And such solution is exactly
equivalent to the matrix obtained by  Jimbo's constructions. In
the present situation (\ref{fs}) there is a point $u_0$ at which
$\check{R}^{r\;
r}(u_0)={I}-P^1=\sum_{r'=3}^{2r-1}P^{r'}$, which can
produce $(r^2-1)\times(r^2-1)$-dimensional $R$-matrices satisfying the YBE defined on the composite spaces.
As these solutions can be interesting in the context of the integrable
models which describe interactions between different kind of
spins, we think it is worthy to obtain the exact form of such
matrices. Of course the construction of the intertwiner matrices
on reducible spaces has also a mathematical interest. Especially
 we shall investigate the series of the homogeneous solutions
defined on the spaces $U^{r^2-1}\otimes U^{r^2-1}$, as well as
 the series of the inhomogeneous descendant
solutions, which can be treated as the "extended" versions of the ordinary Lax
operators.

The discussion hereinafter is proper for both of
  the symmetries of quantum algebra $sl_q(2)$ and super-algebra $osp_q(1|2)$, only
  for the second case the grading of the states must be taken into account. Particularly the tensor product
  must be replaced by the graded tensor product, and the YBE, when $R$-matrices are written in non check formulation, would contain additional signs conditioned by the parities (e.g. see \cite{SHKKH}). For clarity, in the next sections when  concretization will be needed
  we shall use the terminology of the $sl_q(2)$-algebra (basic definitions are brought in the Appendix),
 but obviously the extension to the case of the quantum
superalgebra  $osp_q(1|2)$ is straightforward.

 The descendant solution on the product $V^{r_1}\otimes
 V^{r_2}\otimes V^{r_3} \otimes V^{r_4}$ can be presented by the following product
 of the R-matrices (for this case $r_1=r_2=r_3=r_4=r$)
\bea \check{R}_{12\;34}(u)=\check{R}_{12}(u_0)
\check{R}_{34}(u_0)
\!\Big[
\!\check{R}_{23}(u)\check{R}_{12}(u-\!u_0)
\check{R}_{34}(u-\!u_0)\check{R}_{23}(u-\!2u_0)
\Big]\!
\check{R}_{12}(u_0)\check{R}_{34}(u_0). \label{ru1234}\ena
Recall that every $\check{R}_{ij}(u)$-matrix has the following
decomposition $\check{R}_{ij}(u)={I}_{ij}+f(u)P_{ij}^1$ and
the point $u_0$ is fixed from the equation
$f(u_0)=-1$. % ($u_0=\mathrm{arctanh}{-x}$).

Further the following notations are used $f_{kn}=f(u_{kn})$, where
$u_{kn}\equiv u_k-u_n$. Here to each state, indexed by $i,\; j$ (lines in the graphical representations, Figures 1,2) of the $\check{R}_{ij}(u_{ij})$-matrix there are attached 'velocities' $u_i,\; u_j$. From the figure 2
it is seen, that the 'flows' of the velocities correspond to the 'flows' of the indexes for the non-check
 matrices $R_{ij}=P \check{R}_{ij}$. And we obtain the
following relations among the spectral parameters $u_{14}=u$,
$u_{12}=u_{34}=u_0$, $u_{24}=u_{13}=u-u_0$, $u_{23}=u-2 u_0$. The proof, that the matrix (\ref{ru1234}) satisfies the YBE, can be done just by successively using the YBE on the  matrices $\check{R}_{ij}(u)$. And of course, the operator $\check{R}_{12\;34}(u)$ has the invariance of the corresponding quantum (super-)agebra, as it is a product of the centralizer operators defined on the space $\bigotimes^4_{i=1} V_i^r$.

Expanding the product of the operators in the big parenthesis into
the sum of projection operators, and taking into account that the
terms, which are equivalent to $P^{1}_{12}\cdot\prod P $,
$P^{1}_{34}\cdot\prod P$ or $\prod P \cdot P^{1}_{12}$, $\prod
P\cdot P^{1}_{34}$,  vanish after multiplying by
$({I}-P^{1}_{12})({I}-P^{1}_{34})$, we come to the
expression
\bea\nn
\check{R}_{12\;34}(u)=({I}-P^{1}_{12})({I}-P^{1}_{34})\times\\
\Big({I}+[f_{23}+f_{14}+f_{23}f_{14}]P^{1}_{23}+f_{14}f_{23}P^1_{23}[f_{24}P^1_{12}+f_{13}P^1_{34}+
f_{24}f_{13}P^1_{12}P^1_{34}]P^1_{23}
\Big)\times\;\;\;\\({I}-P^{1}_{12})({I}-P^{1}_{34})\nn
 \ena

In the course of the calculations done in the previous section we
have obtained that the following relations hold
%In terms of the  functions $f_{kn}$ the functional relation
%(\ref{eq-f}) can be written as
(we denote here  $\mathcal{X}\equiv\mathcal{X}(j_0)=C\left(^{j_0\;
j_0\; 0}_{t\;-t\;\; 0}\right)\bar{C}\left(^{j_0\; j_0\;
0}_{t\;-t\;\; 0}\right)C\left(^{j_0\; j_0\; 0}_{-t\;t\;\;
0}\right)\bar{C}\left(^{j_0\; j_0\; 0}_{-t\;t\;\; 0}\right)$)
\bea
P^1_{12}P^1_{23}P^1_{12}=\mathcal{X}P^1_{12},\;\;\;P^1_{23}P^1_{12}P^1_{23}=\mathcal{X}
P^1_{23},\;\;\; P^1_{23}P^1_{34}P^1_{23}=\mathcal{X}P^1_{23}.\ena
And in the same way, acting on the vector  $v_k\otimes v_p\otimes
v_i\otimes v_j$ of the space $V_1\otimes V_2\otimes V_3\otimes
V_4$, we find out  that
\bea
P^1_{23}P^1_{12}P^1_{34}P^1_{23}=\mathcal{X}P^1_{14}P^1_{23}.\ena
Thus we have
\bea %\label{r1234}
\check{R}_{12\;34}^{(r^2-1)(r^2-1)}(u)=({I}-P^{1}_{12})({I}-P^{1}_{34})\times\nn\\
\Big({I}+[f_{23}+f_{14}+f_{23}f_{14}+\mathcal{X}f_{14}f_{23}(f_{24}+f_{13})]P^1_{23}+
\mathcal{X}f_{14}f_{23}f_{24}f_{13}P^1_{14}P^1_{23}
\Big)\times\label{r1234}\;\\({I}-P^{1}_{12})({I}-P^{1}_{34}).\nn
 \ena

 The functional dependence of these matrices for all values of $r$ coincides with the one presented in (\ref{r31}), which corresponds to the descendant matrix of the fundamental irrep with $r=2$.
  The descendant matrix for the case with
$r=3$, which corresponds to the fundamental irrep of $osp_q(1|2)$, is defined on the product of the spaces $(V^3\oplus V^5)\otimes (V^3\oplus V^5)$. And the invariant operators in the expansion
\bea
\check{R}^{8\; 8}(u)=q^u \check{R}^{+,8}+ \check{R}^{0,8}+q^{-u} \check{R}^{-,8},
\ena
consist of the linear superpositions of the following projection operators:
\bea
\check{R}^{\pm/0,8}=\sum_{n,i,j} \mathbf{r}^n_{i,j} P^n_{i,j}, \qquad n=1,3,5,7,9,
\ena
where by $P^n_{i,j}$ the projectors are denoted, which act on the irreps with dimension $n$ as follows -  $P^n_{i,j} V^n_i=V^n_j$. Note, that in the case  of the tensor product of the reducible states (which is true when $r>2$)  in the decomposition  there are few irreps with the same dimensions, and only the  irrep with the maximal dimension has multiplicity 1,
\bea
 (V^3\oplus V^5)\otimes (V^3\oplus V^5)=\bigoplus_{i=1}^2 V^1_i\oplus \bigoplus_{i=1}^4 V^3_i\oplus\bigoplus_{i=1
 }^4 V^5_i\oplus\bigoplus_{i=1}^3 V^7_i \oplus V^9_i.
 \ena
\section{Integrable model on  1D chain $\bigotimes V^{(r^2-1)}$}
 \addtocounter{section}{0}\setcounter{equation}{0}

As it is known,  the spectral parameter dependent Yang-Baxter equations ensure the
integrability of $1+1$ quantum spin models constructed via the
R-matrices  being solutions to YBE. And according to the ideas of the Algebraic Bethe Ansazt the
Hamiltonian operators of the corresponding models are defined by
means of the first order logarithmic derivatives of the transfer
matrices \cite{FTSZ1,baxter,KBI}, $\tau(u)=tr_a \prod_i R_{ai}(u)$ at the point, where $\check{R}$-matrix coincides with unity operator
 %$\check{R}(u_0)=I$
  (this ensures the locality). The index $a$ denotes an
auxiliary space, and the index $i$ - the space situated on the $i$-th site
of a chain.
We are interested here in the chain models connected with the
obtained series of the $R^{(r^2-1)(r^2-1)}$-matrices, for which the quantum
spaces defined on each sites are the states %$r^2-1$-dimensional states
$(V^r\otimes V^r-V^1)=V^{2r-1}\oplus V^{2r-3}\oplus\cdots V^3$.
From the operator form of the matrix (\ref{r1234}) we can check
that at $u=u_0$ the operator $\check{R}_{1234}$ is
$({I}-P^{1}_{12})({I}-P^{1}_{34})$, which equals to
 unity operator in the considered
$(r^2-1)\times(r^2-1)$-dimensional space. This follows from the
functional dependence of the expressions in (\ref{r1234})
\bea
f_{23}+f_{14}+f_{23}f_{14}+\mathcal{X}f_{14}f_{23}(f_{24}+f_{13})=
f(u_{14}+u_{23})(1-\mathcal{X}f_{23}f_{14})+
\mathcal{X}f_{14}f_{23}(f_{24}+f_{13})=\nn\\ { }
f(2u-2u_0)(1-\mathcal{X}f(u)f(u-2u_0))+2\mathcal{X}f(u)f(u-2u_0)f(u-u_0),\qquad\ena
and
\bea
\mathcal{X}f_{14}f_{23}f_{24}f_{13}=\mathcal{X}f(u)f(u-2u_0)[f(u-u_0)]^2,\ena
 and also from the observations, that the
first order series expansion of the function $f(u)$ near the point
$u=0$ equals to
 $f(u)\approx f_0 u$,
 meanwhile  %one can find,
   the point $(-u_0)$
 is  singular, as $f(-u_0+u)\approx \frac{1}{u(1/f_0-f_0/4)}$. It means that
at the point $u_0$ the following expansion is true
\bea f(u)f(u-2u_0)[f(u-u_0)]^2=-\frac{(u-u_0)}{f_0^{-2}
 (1/f_0-f_0/4)}+O(u-u_0)^2.
 \ena
  So we obtain the following expansion near $u_0$,
  \bea\label{r-expan}
  \check{R}_{1234}(u)\!\approx \!
(I\!-\!P^{1}_{12})(I\!-\!P^{1}_{34})\Big[I\!\!+\!
2(u\!-\!u_0)f_0
P^1_{23}\!-\!\mathcal{X}\frac{(u\!-\!u_0)}{f_0^{-2}
 (\frac{1}{f_0}\!-\!\frac{f_0}{4})}P^1_{23}P^1_{14}\Big](I\!-\!P^{1}_{12})
 (I\!-\!P^{1}_{34}).\ena

 And hence,
we can formulate the corresponding  quantum 1d Hamiltonian %
 defined on a cyclic chain having on each site $r^2-1$
dimensional vector space (superposition of the spin states) with the following nearest neighborhood
interactions arising from the first order expansion of the
$R$-matrix (\ref{r-expan})
\bea H=f_0\sum_{i} \left({\bar{P}}_{i\; i+1}+\bar{{\mathcal{X}}}
{\hat{P}}_{i\;i+1}\right).\label{h}\ena
Here the notation  $\bar{{\mathcal{X}}}=\mathcal{X}\frac{2
f_0^{2}}{
 (4\!-\!{f_0}^{2})}$ is used, and the nearest-neighborhood interactions
are described by the operators ${\bar{P}}_{i\; i+1},\;
{{\hat{P}}}_{i\;i+1}$. Provided that the lattice has double
substructure, $i=\{2s,2s+1\}$, i.e. $\mathbb{V}_i=V_{2s}^r\otimes
V_{2s+1}^r$,  the following relations take place \\
${\bar{P}}_{i\; i+1}=\left[(I\!-\!P^{1}_{2s,\;
2s+1})(I\!-\!P^{1}_{2s+2 \; 2s+3})\right] P^1_{2s+1 \;
2s+2}\left[(I\!-\!P^{1}_{2s,\;
2s+1})(I\!-\!P^{1}_{2s+2 \; 2s+3})\right]$ and ${{\hat{P}}}_{i\;
i+1}=\left[(I\!-\!P^{1}_{2s,\;
2s+1})(I\!-\!P^{1}_{2s+2 \; 2s+3})\right] P^1_{2s+1 \;
2s+2}P^1_{2s \; 2s+3}\left[(I\!-\!P^{1}_{2s,\;
2s+1})(I\!-\!P^{1}_{2s+2 \; 2s+3})\right]$.
 The internal part of the Hamiltonian operator describes spin interactions on two sub-chains,
 which can be schematically presented as
  $\sum_s(\mathcal{H}_{2s\; 2s+1}+ \mathcal{H}_{2s-1\; 2s+2})$.
   The existence of the external
 projection operators $\left[(I\!-\!P^{1}_{2s,\;
2s+1})({I}\!-\!P^{1}_{2s+2 \; 2s+3})\right]$ however indicates that there are mixed interactions between four neighboring
 spins
  on the sub-chains positions.

Now let us explore whether
this Hamiltonian is as an operator describing superposition of the pure spin-spin
interactions between the different spins defined at $\mathbb{V}_i$ and $\mathbb{V}_{i+1}$. A task is to ascertain the possibility of  decomposition of the
  operators $\bar{P}_{i\;i+1}$ and $\hat{P}_{i\; i+1}$ in terms of the
 algebra invariant polynomials of the
 spin operators defined on the spaces $\mathbb{V}_i$ and $\mathbb{V}_{i+1}$.
 The expansion of the Hamiltonian
 operators in terms of the algebra generators at the spaces $V_{2s}$ and $V_{2s+1}$ is obvious
 just by construction.

 The question about the spin structure of  $H_{i\;i+1}$ we can achieve  either by the quantum
$6j$-symbols or directly by the Clebsh-Gordan coefficients \cite{kr,WinSM}.  A
brief description how to use the quantum $6j$-symbols  for
obtaining the decomposition is brought in the Appendix A3. From the
analysis of the dimensions in the expansions done therein, it
follows, that actually, we must clarify whether the following
decompositions - $(I-P^1_{12})(I-P^1_{34})P^1_{23}
(I-P^1_{12})(I-P^1_{34})=\sum_{k=1}^{2r-1}\sum_{r',r''}
a_k^{r'\;r''} A^{r'}_{12}A^{r''}_{34}P^{k}_{1234}$,
$\;(I-P^1_{12})(I-P^1_{34})P^1_{23}
P^1_{14}(I-P^1_{12})(I-P^1_{34})=\sum_{r'}a^{r'}
A^{r'}_{12}A^{r'}_{34}P^{1}_{1234}$, are valid, where  $A^{r}$ denotes
 some combinations of algebra  operators  defined on the spaces $\mathbb{V}_{12}$ and $\mathbb{V}_{34}$. It is known that the quantum $6j-$symbols are expressed by the
sums of the quartic products of the Clebsh-Gordan coefficients. And here  we prefer operate immediately with
the Clebsh-Gordan coefficients  as in the previous sections.

\subsection{%
 The structure of the Hamiltonian  operator
 }

Here we use for the
orthogonalized vector states $v^r_k$  the "ket", "bra" notations, $|j,k\rangle$, $r=2j+1$.
The unity operator defined on the space $V^r$ can be expressed as
$I^r=\sum_k \frac{|j,k\rangle \langle j,k|}{\langle
j,k|j,k\rangle}$. The projector operator $P_{r_1
r_2}^{r_0}$ acting on the tensor product $V^{r_1}\otimes V^{r_2}$
and distinguishing the space $V^{r_0}$, $r_0=2j_0+1$, we can write
in this way, by using the formula (\ref{v12})
\bea
 P_{r_1 r_2}^{r_0}=\sum_{i=-j_0}^{j_0} \frac{|j_0,{i}\rangle \langle
j_0,{i}|}{\langle
j_0,i|j_0,i\rangle}=\\\nn\sum_{i=-j_0}^{j_0}
\sum_{i_1=-j_1}^{j_1}\sum_{i'_1=-j_1}^{j_1}C\left(^{j_1 j_2
j_0}_{i_1\; i-i_1\; i}\right)C\left(^{j_1 j_2 j_0}_{i'_1\;
i-i'_1\; i}\right) \frac{|{j_1},{i_1}\rangle
|{j_2},{i-i_1}\rangle \langle
{j_2},{i-i'_1}|\langle {j_1},{i'_1}|}{\langle j_0,{i_0}|j_0,{i_0}\rangle}. \ena

The unity operator $I^{r_1\times r_2}=\sum_{r_0=|r_1-r_2|+1}^{(r_1+r_2)-1}P_{r_1 r_2}^{r_0}$
defined on the space $V^{r_1}\otimes V^{r_2}$ can be written as
\bea I^{r_1\times r_2}=%
\sum_{i_1=-j_1}^{j_1}\sum_{i_2=-j_2}^{j_2}
\frac{|{j_1},{i_1}\rangle |{j_2},{i_2}\rangle\langle {j_2},{i_2}|\langle
{j_1},{i_1}|}{\langle
{j_1},{i_1}|{j_1},{i_1}\rangle \langle
{j_2},{i_2}|{j_2},{i_2}\rangle}. \ena

For the orthosymplectic algebra one must take into account the grading of the vectors, and  appropriate signs would appear in the above formulas. The vector states $v_{i_k}^{r_k}$ can be chosen to be normalized,
then the formulas would be more compact. The first term of the
Hamiltonian operator corresponding to the cell $V_1\otimes V_2\otimes V_3\otimes V_4$ can be presented as the following,
taking into account that $j_1=j_2=j_3=j_4\equiv j$,
\bea& ({I}\!-\!P^{1}_{12})({I}\!-\!P^{1}_{34})\left[
P^1_{23}\right]({I}\!-\!P^{1}_{12})
 ({I}\!-\!P^{1}_{34})=&\label{p23}\\&
 \sum_{j_{1234}=j'_{1234}}^{2j}\;\;\;
 \sum_{j_{12},j_{34},j'_{12},j'_{34}=1}^{2j}\;\;\sum_{i,p,k,k'=-j}^j
 \nn C\left(^{j\;\;\; j \;0}_{k\; -k\;
0}\right) C\left(^{j\;\;\;j\; 0}_{k'\; -k'\; 0}\right)%\times&\\&
 \bar{C}\left(^{j\; j\; \;j_{12}}_{p\; k\;
p+k}\right)\times&\\&\bar{C}\left(^{j\; j\; \;j'_{12}}_{p\; k'\;
p+k'}\right)%\times\nn\\
\bar{C}\left(^{\;j\;\; \;\;j\;\quad
j_{34}}_{-k\; i-p\;i-p-k}\right)\bar{C}\left(^{\;j\;\;\;\;\;
j\;\quad j'_{34}}_{-k'\; i-p\;
i-p-k'}\right)\bar{C}\left(^{\;j_{12}\;\quad\;
j_{34}\;\;j_{1234}}_{p+k\; i-p-k\;\;\;
i}\right)\bar{C}\left(^{\;j'_{12}\;\quad\; j'_{34}\;\;
j'_{1234}}_{p+k'\; i-p-k'\;\;\; i}\right)\times&\nn\\&{|{j_{1234}},{i}\rangle_{j^{ }_{12},j{ }_{34}}\;\; { }_{{j'_{12},j'_{34}}}\!
\langle {j'_{1234}},{i}|}{ }.&\nn\ena
Correspondingly, the second term of the Hamiltonian operator will be
\bea& ({I}\!-\!P^{1}_{12})({I}\!-\!P^{1}_{34})\Big[
P^1_{23}P^1_{14}\Big]({I}\!-\!P^{1}_{12})
 ({I}\!-\!P^{1}_{34})=%=&\label{p23p14}\\&
 \sum_{j_{12}=j_{34}=1,j'_{12}=j'_{34}=1}^{2j}\;\;\sum_{p,p',k,k'=-j}^j
 &\label{p23p14}\\&\nn C\left(^{j\;\; j \;0}_{k\; -k\;
0}\right) C\left(^{j\;\; j\; 0}_{k'\; -k'\; 0}\right)
C\left(^{j\;\; j \;0}_{p\; -p\; 0}\right) C\left(^{j\;\; j\;
0}_{p'\; -p'\; 0}\right)
 \bar{C}\left(^{j\; j\;\; j_{12}}_{p\; k\;
p+k}\right)\bar{C}\left(^{j\; j\; \;\; j'_{12}}_{p'\; k'\;
p'+k'}\right)\times\nn&\\&\bar{C}\left(^{\;j\;\; \quad j\;\quad
j_{34}}_{-k\; -p\;- p-k}\right)\bar{C}\left(^{\;j\;\quad\;
j\;\quad j'_{34}}_{-k'\; -p'\;
-p'-k'}\right)\bar{C}\left(^{j_{12}\;\quad j_{34}\;\;\; 0}_{p+k\;
-p-k\; 0}\right)\bar{C}\left(^{j'_{12}\;\quad j'_{34}\;\;\;
0}_{p+k'\; -p-k'\; 0}\right){|1,0\rangle_{j^{
}_{12},j^{ }_{34}}\;\; { }_{{j'_{12},j'_{34}}}\!
\langle{1},{0}|}%{(\langle v_i^{1}|v_i^{1}\rangle)^2}
.&\nn \ena

The obtained operators constitute superpositions of the projectors
$(P^{r_0})_{a,b}:(V^{r_0})_a\to (V^{r_0})_b$, which are algebra
invariant operators mapping the different spaces $(V^{r_0})_{a,b}$
with the same spin $j_0=(r_0-1)/2$ ($\equiv j_{1234}=j'_{1234}$ in (\ref{p23})) arising in the fusion of
$V^r\otimes V^r\otimes V^r\otimes V^r$ one to other. For the
expression in Eq. (\ref{p23}) $r_0=1,...,2r-1$, and for the case of
Eq.(\ref{p23p14}) - $r_0=1$).

Now let us turn to the question arisen just before this subsection. We can see that the mentioned
 decomposition in general does not take place. In the case of the
$H$-operator (\ref{p23}) the action on the space $\mathbb{V}_{12}\otimes
\mathbb{V}_{34}$ is performed by the linear superpositions of the
projectors $\sim \sum\prod C()\bar{C}()\left[|v_p^{r^{ }_{12}}\rangle\langle
v_{p'}^{r'_{12}}|\right]\left[|v_{i-p}^{r^{ }_{34}}\rangle\langle
v_{i-p'}^{r'_{34}}|\right]$ (emerging at the intermediate stage of the decompositions in Eq. (\ref{p23}), in $\prod C()\bar{C}()$ only the first six Clebsh-Gordan coefficients from Eq. (\ref{p23}) are included), which means that the action of the Hamiltonian changes
not only the values of the algebra operator $h$ (spin's projectors), but also the kind
of the irreps  belonging to the spaces $\mathbb{V}_{12}=V_1\otimes V_2$ and
$\mathbb{V}_{34}=V_3\otimes V_4$. In the  chain $\bigotimes_i \mathbb{V}_{i}$,
where $i=\{2k,2k+1\}$, the action of the corresponding part of the
Hamiltonian operator $H=\sum_i H_{i,i+1}$ can be schematically
presented by the formula
\bea H_{i,i+1}\simeq \sum_{r,r'}a_{r_i,r_{i+1}}^{r'_i,r'_{i+1}}
\mathcal{J}_{r_i}^{r'_i}\mathcal{J}_{r_{i+1}}^{r'_{i+1}}, \label{hj}
\ena
 here we have used this formal notation - $\mathcal{J}_{r}^{r'}=|v^r\rangle\langle v^{r'}|$, where the
indexes of the spin projections are omitted.
 In case of (\ref{p23p14}) $r_i=r_{i+1},\;\;\;r'_i=r'_{i+1}$, as $r_{1234}=1$ and $r_{12}=r_{34}$,
$r'_{12}=r'_{34}$.
 And clearly, the
operators $\mathcal{J}_{r}^{r'}$ in general ($r\neq r'$) are not expressed by the algebra
generators defined on the states $\mathbb{V}_i$.

Detailed expressions and the study of such quite large Hamiltonian operators for specific cases we  purpose to do in subsequent work.

 The actual spaces %
 on which the Hamiltonian operator (\ref{h}) is acting, are the truncated products, and in the next discussions we shall use  new notations $U^{\mathcal{R}}$ for denoting such $\mathcal{R}$-dimensional composed spaces. Particularly,   by the action of the projectors the following product of the irreps  $\mathbb{V}_i={V^r}_{2s_i}\otimes V^r_{2s_i+1}$  turns into   $r^2-1$-dimensional state - $(I-P^1_{2s_i 2s_i+1})\mathbb{V}_i(I-P^1_{2s_i 2s_i+1})\equiv
 {V^r}_{2s_i}\otimes V^r_{2s_i+1}-V^1_{i}$, and will be denoted as $U^{{r^2-1}}$.

\section
{Centralizers and reducible representations.}

 In fact in (\ref{hj}) we deal with a centralizer operator defined
on the tensor product of two mixed states $U^{r^2-1}\otimes U^{r^2-1}$. Here we intend by a straightforward construction to reveal the structure of such operators. Let us write down the conditions which
the algebra relations put on the centralizers defined on the product $U\otimes U$, with composite representations spaces $U$, and then the extension to general case $U\otimes U\otimes ...\otimes U$ can be done by similar calculations. Let $U$ consists of some set of irreducible representations: $U=\bigoplus_{r_k} V^{r_k}$. The thorough formulation of the operators $\mathcal{J}_{r}^{r'}$ could be done by means of  the ortho-normalized elementary operators
\bea
\mathcal{J}_{j,i}^{j',i'}\equiv| j',i'\rangle| \langle j,i|,\quad \mathcal{J}_{\bar{j},\bar{i}}^{\bar{j}',\bar{i}'}\mathcal{J}_{j,i}^{j',i'}=
\delta_{j'}^{\bar{j}}
\delta_{i'}^{\bar{i}}\mathcal{J}_{j,i}^{\bar{j}',\bar{i}'}.
\ena
 If in $U$ there are more than one copies of the irreps with the given same spin-$j$, one must
 add an additional index for differentiating them, e.g. - $j_a$. Each linear operator $\mathbf{a}$, evaluating
 in $U$ can be presented as a superposition
\bea \mathbf{a}=\sum_{j_a,j'_b,i_a,i'_b} a_{j_a,i_a}^{j'_b,i'_b}\mathcal{J}_{j_a,i_a}^{j'_b,i'_b}. \ena
 The algebra generators on this basis can be presented as (the coefficients $\beta,\;h,\;\gamma$ below
 denote the usual matrix elements of the corresponding operators, for the quantum super-algebra $osp_q(1|2)$
 see the Appendix A.1)
 \bea e=\sum_{j_a,i_a} \beta^{j_a}_{i_a}\mathcal{J}_{j_a,i_a}^{j_a,i_a+1},\quad
 f=\sum_{j_a,i_a} \gamma^{j_a}_{i_a}\mathcal{J}_{j_a,i_a}^{j_a,i_a-1},\quad
 h=\sum_{j_a,i_a} h^{j_a}_{i_a}\mathcal{J}_{j_a,i_a}^{j_a,i_a}.\ena
 The matrix elements of the generators just by definition are the same (up to some elementary transformations, admissible by the algebra relations, see e.g. \cite{SHKKH}) for the irreps with the same spin $\beta^{j_a}_{i_a}\equiv\beta^{j}_{i}, ...$.  Every element of the center defined on $U$ consists of these elementary projection operators
 \bea P_{j_a}^{j_b}=\sum_{i=-j}^j \mathcal{J}_{j_a,i}^{j_b,i}, \qquad \forall j,\;\;\forall a,\; b.\ena
 The quadratic Casimir operator is just the sum $\mathbf{c}=\sum_{j_a}\mathbf{c}_{j}P_{j_a}^{j_a}$. The centralizer operators $c$ defined on the tensor products of $n$-copies of the mixed states $U$,
 $U\otimes U\otimes \cdots\otimes U$ have more rich structure. The sufficient and necessary conditions for them can be obtained straightly from the equations $[\Delta(\Delta\otimes\cdots(\Delta\otimes I))[g],c]=0$, presenting the commutation of the operators $c$ with the algebra generators defined on the tensor product of the representations by means of the associative co-product operation. In general one can write any linear operator defined in $\bigotimes^n U$ as
 \be\mathbf{a}=\sum_{j_{ak},i_{ak},j'_{bk},i'_{bk}} a_{j_{a1},i_{a1};j_{a2},i_{a2};...;j_{an},i_{an}}^{j'_{b1},i'_{b1};j'_{b2},i'_{b2};...;j'_{bn},i'_{bn}}
 \mathcal{J}_{j_{a1},i_{a1}}^{j'_{b1},i'_{b1}}\otimes \mathcal{J}_{j_{a2},i_{a2}}^{j'_{b2},i'_{b2}}\otimes\cdots \mathcal{J}_{j_{an},i_{an}}^{j'_{bn},i'_{bn}},\ee
 or briefly $\mathbf{a}=\sum_{\{j;i\}} a_{\{j_{a},i_{a}\}}^{\{j'_{b},i'_{b}\}}
 \bigotimes^n \mathcal{J}_{j_{a},i_{a}}^{j'_{b},i'_{b}}$. For the generic  centralizer operators $\mathbf{c}$ in the simplest case $n=2$ (which is enough to consider if we are interested with the nearest-neighbourhood interaction Hamiltonians) the
 thorough calculations give the following relations on the coefficients $c_{j_{a1},i_{a1};j_{a2},i_{a2}}^{j'_{b1},i'_{b1};j'_{b2},i'_{b2}}$, ensuring the commutation of $\mathbf{c}=\sum_{\{j;i\}} c_{\{j_{a},i_{a}\}}^{\{j'_{b},i'_{b}\}}
 \bigotimes^n \mathcal{J}_{j_{a},i_{a}}^{j'_{b},i'_{b}}$, $n=2$, with the algebra generators,
\bea\label{g,c}
&\!\!\!\!\!\!\!\!\!\!i_{a1}+i_{a2}=i'_{b1}+i'_{b2},&\\&\!\!\!\!\!\!\!\!\!\!
\beta_{i'_1-1}^{j'_1} c_{j_{a1},i_{a1}\;\;;j_{a2},i_{a2}\;\;\;}^{j'_{b1},i'_{b1}-1;j'_{b2},i'_{b2}+1}+q^{h^{j'_1}_{i'_1}} \beta_{i'_2}^{j'_2} c_{j_{a1},i_{a1};j_{a2},i_{a2}}^{j'_{b1},i'_{b1};j'_{b2},i'_{b2}}=\beta_{i_1}^{j_1} c_{j_{a1},i_{a1}+1;j_{a2},i_{a2}\;\;}^{j'_{b1},i'_{b1}\;\;;j'_{b2},i'_{b2}+1}+q^{h^{j_1}_{i_1}} \beta_{i_2}^{j_2} c_{j_{a1},i_{a1};j_{a2},i_{a2}+1}^{j'_{b1},i'_{b1};j'_{b2},i'_{b2}+1},\nn&\\&\!\!\!\!\!\!\!\!\!\!\nn
q^{-h^{j'_2}_{i'_2-1}}\gamma_{i'_1+1}^{j'_1} c_{j_{a1},i_{a1}\;\;;j_{a2},i_{a2}\;\;\;}^{j'_{b1},i'_{b1}+1;j'_{b2},i'_{b2}-1}+ \gamma_{i'_2}^{j'_2} c_{j_{a1},i_{a1};j_{a2},i_{a2}}^{j'_{b1},i'_{b1};j'_{b2},i'_{b2}}=q^{-h^{j_2}_{i_2}} \gamma_{i_1}^{j_1} c_{j_{a1},i_{a1}-1;j_{a2},i_{a2}\;\;}^{j'_{b1},i'_{b1}\;\;;j'_{b2},i'_{b2}-1}+ \gamma_{i_2}^{j_2} c_{j_{a1},i_{a1};j_{a2},i_{a2}-1}^{j'_{b1},i'_{b1};j'_{b2},i'_{b2}-1}.&
\ena
Here there is taken into account that $h_i^j\equiv i$. For the case of the quantum super-algebra $osp_q(1|2)$ one must take into account the parities of the states, the graded character of the tensor products and that $h_i^j=i+\imath\; constant$ for the even dimensional irreps.

 To obtain the corresponding relations for the general case with arbitrary $n$ is an obvious task, which brings to the evident extension of the Eqs.(\ref{g,c}), with $n$ summands at the r.h.s. and l.h.s. of the equations.
\bea&\sum_{k=1}^n i_{ak}=\sum_{k=1}^n i'_{bk},&\label{hn}\\\nn%\label{bn}
& \sum_k^n q^{\sum_{p<k}h^{j'_p}_{i'_p}}
\beta_{i'_k-1}^{j'_k} c_{j_{a1},i_{a1};...\;j_{ak},\;i_{ak}\;;...j_{an},i_{an}\;\;\;}^{j'_{b1},i'_{b1};...j'_{bk},i'_{bk}-1;...j'_{bn},i'_{bn}+1}
=\sum_k^n q^{\sum_{p<k} h^{j_p}_{i_p}} \beta_{i_p}^{j_p} c_{j_{a1},i_{a1};...j_{ak},i_{ak}+1;...j_{an},i_{an}\;\;}^{j'_{b1},i'_{b1};...\;\;j'_{bk},i'_{bk};...
j'_{bn},i'_{bn}+1},&\\&\!\!\!\!\!\!\!\!\!\!
\sum_k^n q^{-\sum_{p\geq k} h^{j'_p}_{i'_p-1}}\gamma_{i'_k+1}^{j'_k} c_{j_{a1},i_{a1};...\;j_{ak},\;i_{ak}\;;...j_{an},i_{an}\;\;\;}^{j'_{b1},i'_{b1};...
j'_{bk},i'_{bk}+1;...j'_{bn},i'_{bn}-1}=\sum_{k}^n q^{-\sum_{p\geq k}h^{j_p}_{i_p}} \gamma_{i_p}^{j_p}c_{j_{a1},i_{a1};...j_{ap},i_{ap}-1;...j_{an},i_{an}\;\;}^{j'_{b1},i'_{b1};,,,j'_{ap},\;i'_{ap};
...\;\;j'_{bn},i'_{bn}-1}.&\nn%\label{gn}
\ena
Particularly the first equation  in Eqs.(\ref{hn}) ensures the conservation of the spin projection.  The next equations put the relations on the coefficients
 $c_{\{j_{a},i_{a}\}}^{\{j'_{b},i'_{b}\}}$. Note, that for the $n$-th term in the sum of the l.h.s of the second equation in (\ref{hn}) one must take $\{i'_{bn}+1\to i'_{bn}\}$, and correspondingly for the similar term of the third equation of (\ref{hn}) - $\{i'_{bn}-1\to i'_{bn}\}$.

\section{Extended Lax operators}
 \addtocounter{section}{0}\setcounter{equation}{0}

 The Hecke type matrices $R^{rr}(u)$ %
  do not allow generalizations to
 the inhomogeneous $R^{rr'}(u)$ acting on $V^{r}\otimes V^{r'}$
  with $V^{r'}$ being  an irrep, so that
 $R^{rr}R^{rr'}R^{rr'}=R^{rr'}R^{rr'}R^{rr}$, besides of the case  $r=2$ with the fundamental irrep $V^2$,
 which gives standard
 universal Lax
 operator $L$, obeying the quantum YBE (see for the references in \cite{SHKKH}, where the corresponding operator is constructed for the case of $osp_q(1|2)$)
\bea RLL=LLR.\label{exL}\ena
  However by  descendant procedure
 we can get algebra invariant matrices
 $R^{r\mathcal{R}}:V^{r}\otimes U^{\mathcal{R}}$ satisfying to YBE with $U^{\mathcal{R}}$ to be a
 composite representation. The simplest $U^{\mathcal{R}}$ %composite representation
 has been discussed in the previous sections.  Applying fusion method further
 one can find descendant $R$ operators defined on the representations larger than $V^r\otimes V^r-V^1$.
 And by means of these operators one can construct
 new quantum integrable models on the 1d chains with the action space
  ${\bigotimes}_i U_i^{\mathcal{R}}$.
  One can expect that
   the corresponding local quantum Hamiltonian operators would describe
 interactions between different spins in a rather entangled way, relying to the discussed example of the series of solvable models
 with homogeneous $R^{r^2-1\;r^2-1}$-matrices.

%%%%%%%%%%%%%%%%%%%%%%%%%%%%%%%%%%%%%%%%%%%%%%%%%%%%%%%%%%%%%%%%%%%
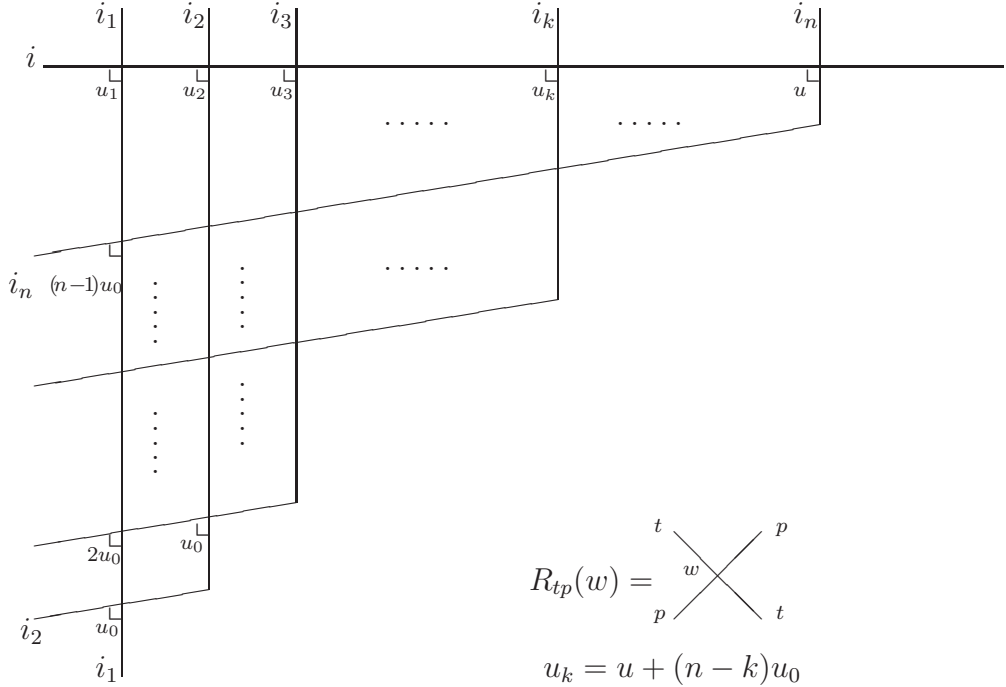
\begin{figure}[t]
\unitlength=11pt
\begin{picture}(100,25)(-1,-1)
\put(3.3,23){\line(1,0){33}}\put(2.7,23){$i$}\put(6,20){\line(0,1){5}}
\put(5.1,24.5){$i_1$}\put(5.1,22){\scriptsize$u_1$}
\put(5.5,22.5){$\llcorner$}\put(9,20){\line(0,1){5}}\put(8.1,24.5){$i_2$}
\put(8.1,22){\scriptsize$u_2$}\put(8.5,22.5){$\llcorner$}
\put(12,20){\line(0,1){5}}\put(11.1,24.5){$i_3$}
\put(11.5,22.5){$\llcorner$}\put(11.1,22){\scriptsize$u_3$}\put(21,20){\line(0,1){5}}
\put(20.1,24.5){$i_k$}\put(20.1,22){\scriptsize$u_k$}\put(20.5,22.5){$\llcorner$}
\put(30,21){\line(0,1){4}}\put(29.1,24.5){$i_n$}\put(29.1,22){\scriptsize$u$}
\put(29.5,22.5){$\llcorner$}
\put(30,21){\line(-6,-1){27}}
\put(20.5,2){$u_k=u+(n-k) u_0$} %%%%%%%%%%%%%%
\put(20,5){${R}_{tp}(w)%({\scriptsize u_{ij}})
=$}\put(25,4){\line(1,1){3}}\put(25,7){\line(1,-1){3}}
\put(24.3,4){\scriptsize$p$}\put(24.3,7){\scriptsize$t$}\put(28.5,4){\scriptsize$t$}
\put(28.5,7){\scriptsize$p$}
\put(25.3,5.5){\scriptsize$w$} %%%%%%%%%%%%%%
 \put(21,15){\line(0,1){5}}
\put(21,15){\line(-6,-1){18}} \put(12,8){\line(0,1){15}}
\put(12,8){\line(-6,-1){9}}\put(9,5){\line(0,1){20}}
\put(9,5){\line(-6,-1){6}}\put(6,2){\line(0,1){23}}\put(5.1,2){$i_1$}
\put(5.5,4){$\llcorner$}\put(5.1,3.5){\scriptsize$u_0$}
\put(5.5,6.5){$\llcorner$}\put(4.8,6){\scriptsize$2u_0$}\put(2.5,3.3){$i_2$}
\put(2.1,15.2){$i_n$}\put(5.5,16.5){$\llcorner$}\put(3.5,15.3){\scriptsize$(\!n\!-\!\!1\!)u_0$}
\put(8.5,6.9){$\llcorner$}\put(8,6.4){\scriptsize$u_0$}
\multiput(23,21)(0.5,0){5}{{.}}\multiput(15,21)(0.5,0){5}{{.}}
\multiput(15,16)(0.5,0){5}{{.}}\multiput(10,14)(0,0.5){5}{{.}}
\multiput(7,13.5)(0,0.5){5}{{.}}\multiput(10,10)(0,0.5){5}{{.}}
\multiput(7,9)(0,0.5){5}{{.}}
\end{picture} \caption{ $R^{r\; {\mathcal{R}_n}}$-matrix ($R_{i\{i_1 ...i_n\}}(u)$)}\label{figrp}
\end{figure}
%%%%%%%%%%%%%%%%%%%%%%%%%%%%%%%%%%%%%%%%%%%%%%%%%%%%%%%%%%%%%%%%%%%%%%%%%%%%%%%%%

So, we can construct the series of the "extended Lax operators"
$L=R^{r\mathcal{R}}$, satisfying (\ref{exL}) with Hecke type $R^{rr}$, and
of course the case of $r=2$ would be the usual Lax operator. In
the same way, as in the previous sections the descendant series
can be constructed by the products of the $R^{rr}(u)$-operators
appropriately fixing the values of the spectral parameters
\cite{j1}. The matrix deduced from the action of the operators
 on the space $V^r\otimes V^r \otimes V^r$ will be
\bea\check{R}^{r\;r^2-1}(u)=\big[I\otimes(I-P^1)\big](\check{R}^{r\;r}(u+u_0)\otimes
I)(I\otimes \check{R}^{r\;r}(u))\big[(I-P^1)\otimes I\big], \ena
defined eventually on $V^r\otimes\Big(V^r \otimes V^r-V^1\Big)$.
Note, that the projection operator $(I-P^1)$ at the right of this
expression could be omitted due to YBE, which actually ensures its
existence. The extension to the space with $n$-product
$V^r\otimes V^r\otimes\ldots\otimes V^r$ we can perform
repeatedly using YBE and truncating by the appropriate projectors,
achieved by taking step by step $u=u_0$ for the right-edge
$R^{rr}(u)$-matrices in the Fig.[\ref{figrp}]. In the resulting
matrix the all spectral parameters are established in accordance
to the summation rule of the additive spectral parameters in YBE,
and, as it was hinted above, only left-hand side projections are taken
into account. So, the matrix $R^{r\times {\mathcal{R}}}(u)$
defined on the truncation of the tensor product
$V^r_i\otimes\left(V^r_{i_1}\otimes\cdots V^r_{i_n}\right)$ can be
written formally as  the following expression:
\bea\label{rnp} {R}^{r\mathcal{R}_n}_{i\{i_1\cdots
i_n\}}(u)=\big[R^{rr}_{i_2 i_1}(u_0)R^{rr}_{i_3
i_1}(2u_0)R^{rr}_{i_3 i_2}(u_0)\cdots R^{rr}_{i_k
i_p}((k-p)u_0)\cdots R^{rr}_{i_n
i_{n-1}}(u_0)\big]\times\\\big[{R}^{rr}_{i
i_1}(u+(n-1)u_0)R^{rr}_{i i_2}(u+(n-2)u_0)\cdots R^{rr}_{i
i_n}(u)\big],\label{rn}\ena
 where the matrices
are presented in "non-check" form, and  the low indexes of the
R-matrices  show the spaces on which the operators act in the
tensor product. The product of the operators in the second row
(\ref{rn}) itself is a solution of the YBE defined on
$V^r_i\otimes\left(V^r_{i_1}\otimes\cdots V^r_{i_n}\right)$, the
expression in (\ref{rnp}) realizes the projection operation
(recall that $\check{R}^{rr}(u_0)=(I-P^1)$). Step by step acting on the
tensor product of the R-matrices these projectors narrow
(restrict) the action space from $r^{n+1}$-dimensional space to
the $r\times\mathcal{R}_n$-dimensional space $V^r\times
U^{\mathcal{R}_n}$, where $U^{\mathcal{R}_n}$ is a reducible
space with dimension $\mathcal{R}_n$, which can be obtained as for
the case of fundamental irrep ($r=2$), from the recurrence
formulas: $\mathcal{R}_0=1,\;\mathcal{R}_1=r,\;\mathcal{R}_2=r^2-1$, ...
$\mathcal{R}_n=r\times\mathcal{R}_{n-1}-\mathcal{R}_{n-2}$. Correspondingly we can retrieve   the
structure of the composite space $U^{\mathcal{R}_n}$: $U^{\mathcal{R}_0}=I,\;U^{\mathcal{R}_1}=V^1,\;...,\;U^{\mathcal{R}_{n+1}}\oplus U^{\mathcal{R}_{n-1}}=V^r\otimes U^{\mathcal{R}_n}$, repeatedly using the fusion rules (\ref{fusion}).
 For the case $r=2$, of course we recover $\mathcal{R}_n=n+1$ and in
 this case $U^{\mathcal{R}_n}=V^{n+1}$ is the $(n+1)$-dimensional irreducible representation - $\frac{n}{2}$-spin irrep. Generally we can refer to the space $U^{\mathcal{R}_{n}}$ as a
 truncated product of the irreps $\otimes^n V^r$.

 As for the explicit formula for this 'extended' Lax operator, from the discussion above
 it follows, that ${R}^{r \mathcal{R}_n}(u)$ possesses  the form
 $P(P^{\mathcal{R}_{n+1}}+P^{\mathcal{R}_{n-1}}+F_n(u)P^{\mathcal{R}_{n-1}})$, where $F_n(u)\approx\prod_k^n f(u+(n-k)u_0)$. The proof also can be done by induction, noting, that the operator  ${R}^{r\;\mathcal{R}_{n+1}}(u)$ differs from ${R}^{r\;\mathcal{R}_{n}}(u)$ by the product of (n+1)-operators $R^{rr}(u+n u_0)\prod_k^{n}R^{rr}((k-1)u_0)$, and the spectral parameter dependent terms in ${R}^{r\;\mathcal{R}_{n+1}}(u)$
  proportional to $\sim f(u+n u_0)$ and $\sim F_n(u)$ are eliminating due to relations coming from YBE, and the action of the projectors $(I-P^1)$, and the only spectral parameter dependent  term which survives is $\sim f(u+n u_0)F_n(u)\Rightarrow F_{n+1}(u)$. So, we can summarize
 \bea
 {R}^{r\mathcal{R}_n}(u)=P(P^{\mathcal{R}_{n+1}}+P^{\mathcal{R}_{n-1}}+f_{r,n,q}\prod_{k=1}^n f(u+(n-k)u_0)P^{\mathcal{R}_{n-1}}). \label{r-inho}
 \ena
The coefficient $f_{r,n,q}$ is conditioned by the action of the mentioned projection operators, as well by the permutation of the spaces, and can be formulated by means of the appropriate set of the CG-coefficients for each case. It is notable to remark, that it is easy to find out the product $\prod_{k=1}^n f(u+(n-k)u_0)$, using %
 the recurrent relations, deduced from the equation (\ref{eq-f})
\be \label{rec}f(u+u_0)=\frac{-1}{1+\chi(j_0)f(u)}.\ee
 This gives for the mentioned product a polynomial of this kind $\frac{(-1)^n f(u)}{1+a_n +b_n f(u)}$.

 Formally these operators keep the form $[P^a+h(u) P^b]$ with two projectors, which  is typical for the Hecke type operators and ensure
 the availability of the Hecke relations on the eigen-vectors' space of $[P^{\mathcal{R}_{n+1}}+P^{\mathcal{R}_{n-1}}]$ which
 plays the role of the unity operator. Only here one must be careful managing with the multiplication or with the inverse operations, as for the inhomogeneous $R$-matrices (\ref{r-inho}) we deal with the transposition (permutation) of the vector states $V^r$ and $U^{\mathcal{R}_n}$.

 For the case of the fundamental irrep, $r=2$,
   fixing $a=\log{q}$ in (\ref{fs}) (note, that there for the case of $sl_q(2)$ we must take into account that $q\to \imath q^{1/2}$) we shall come to $\prod_{k=1}^n f(u+(n-k)u_0)=\frac{(-1-q^2)^n(q^{2u}-1)}{q^{2u}-q^{2n}}$,  which leads to the usual form of the Lax operator for the algebra $sl_q(N)$: $L(u)= q^{u} L^{+}-q^{-u} L^{-}$ \cite{isa}. For the general cases with $r>2$, the corresponding and \textit{similar} expansion of the "extended" Lax operator $R^{r\mathcal{R}}$ is followed by taking into account the presented above polynomial formulae coming from the recurrent relation (\ref{rec}).
%  if the  and so

 All the same results are valid in the non-deformed
 case also, as the corresponding limit ($q\to 1$) is well defined
 (recall merely, that for the ortho-symplectic algebra  at the classical limit only odd dimensional
 irreps are existing).

The one-chain Hamiltonian operators, corresponding to the obtained inhomogeneous matrices (\ref{r-inho}), constructed by means of the transfer matrices, where the auxiliary space is the $V^r$-irrep,  describe  non trivial interactions between different spins, having the structure expressed by the projection operators, similar to the one, discussed in the previous section.
 Strict investigations of such Hamiltonian structures will be carried out in the  future.

 The set of the composed states, fitting to the truncated tensor products of the spin-irreps, can be built for each case separately. For the simple case $n=2$, the normalized (but not orthogonal) states of $U^{r^2-1}$, induced from the initial sublattice,  are determined elementary, using the relation (\ref{12v}):  $|\psi_{i,k}\rangle=\frac{\sum_{j'>0}^{2j}\bar{C}\left(^{j\; j\;\;\;\;\;j'}_{i\;k\; i+k}\right)|j',\;i+k\rangle}{\left[1-\bar{C}\left(^{j\;\; j\;\;\;\;0}_{i\;-i\;\; 0}\right)^2\delta_{(i+k)\;0}\right]^{1/2}}$.

 \addtocounter{section}{0}\setcounter{equation}{0}
   \section{Summary and conclusions}

  Summing up, we can state, that here new type of solutions to Yang-Baxter equations defined on the composite states has been investigated, in particular, the solutions obtained by the fusion method from the  solutions of Hecke type. The Hecke type homogeneous $R^{rr}$-matrix's series has been constructed for the
  quantum super-algebra $osp_q(1|2)$, defined on the
  tensor product of  two $r$-dimensional
  irreducible representations, quite analogous to the corresponding matrix with $sl_q(2)$ symmetry. This pattern ascertains that
  the equivalence of the representation spaces of two algebras
  implies the equivalence of the solutions to YBE, as here the
  important role have the basis operators for the $R$-matrices,
  i.e. the algebra invariant operators - projectors. So, the YBE solutions
  known for the algebra $sl_q(2n)$ must be valid for the
  $osp_q(1|2n)$ quantum super algebra after the appropriate
  changes connected with the gradings and quantum deformation parameter $q$, as there is full correspondence
  between their representations \cite{z}.

   For the Hecke type YBE solutions the corresponding descendant series $R^{r^2-1\; r^2-1}$ have been constructed, which are
  defined on the composite (reducible) $r^2-1$-dimensional states  of
the  $sl_{{q}}(1|2)$ (or $osp_q(1|2)$) algebra. Also descendant inhomogeneous matrices $R^{r \mathcal{R}}$ ("extended" Lax operators), compatible with the mentioned invariant series, have been  suggested,  with definite series of $\mathcal{R}_n$-dimensional composite states  for each $r$-dimensional irrep. Of course, more general "extended" $R$-operators $R^{\mathcal{R}\mathcal{R}'}$ also could be observed, which would be descendant matrices inherited from the obtained ones. All such type of $R$-operators
  produce Hamiltonian operators corresponding to 1d quantum integrable spin
  models describing non-elementary mixed interactions between different kinds of spins situating
  on the sites of the chains.

  %q, roots of unity, special values if q Indecomposable representations.
  We can summarize the results schematically by the following diagram of the series of YBE solutions with the $sl_q(2)$ ($osp_q(1|2)$) symmetry, including the obtained descendant matrices defined on the composite states (and inherited from the Hecke type $R^{rr}$ matrices with $r>2$)  together with the universal $R^{rr'}$ matrices defined on the irreps, which are descendant matrices originated from the matrix on the fundamental irreps $R^{22}$
  $$R^{22}({ }_{universal\;matrix/Hecke\;type}){ }_{------\to}^{descendants} R^{2r}({ }_{universal,\;ordinary\;Lax})\to R^{rr'}({ }_{universal}),$$ $$R^{33}({ }_{Hecke\;type}){ }_{------\to}^{descendants} R^{3\{8,21,...\}}({ }_{Hecke\;type})\to R^{\{8,21,...\}\{8,21,...\}}$$$$\vdots$$
  $$R^{rr}({ }_{Hecke\;type}){ }_{------\to}^{descendants} R^{r\mathcal{R}_n}({ }_{extended\; Lax,\;Hecke\;type})\to R^{\mathcal{R}_n\mathcal{R'}_{n'}}({ }_{extended\; R})$$

   This diagram does not exhaust  all the variety of the YBE solutions for the given symmetries, particularly the solution $R^{33}$ (\ref{r32}) has not been involved here, but it does represent the interrelated series of the solutions %, %having the same matrix with minimal dimension - %$R^{22}$,
     and their descendants.

  \paragraph{Further developments and applications.}   %Hamiltonian operators
One can go further, and together with the descendant series of the Hecke type solutions $R^{rr}$ (such as the "extended" $L$-matrices $R^{r \mathcal{R}}$ or $R$-matrices
$R^{\mathcal{R}\mathcal{R}'}$), which are YBE solutions just by construction, consider any $R$-matrix, defined on the tensor product of the arbitrarily composed sets of the irreps. This can be achieved using the method of the construction of the centralizer operators on the tensor product of the reducible representations, brought in this article, and then
 try to solve the Yang-Baxter equations representing the $R$-matrix in the form of a superposition of the centralizers. Quite equivalently, one can construct all the possible projection operators which exist for the tensor product of the given composite representations, and represent the $R$-matrix as an expansion over these projectors.

Note, that at the exceptional values of the deformation parameter of the quantum group (i.e, when $q$ is a root of unity), the specter of the irreducible representations is restricted,  higher spin irreps are deforming, and new indecomposable representations are arising \cite{qru}, and correspondingly, the fusion rules also are deformed,  but however in this case also the %
solutions of YBE defined on the composed states can be found, properly defining the centralisers or the projection operators  (see \cite{SHKH,shk} and the references therein). As example, at $q^4=1$ the descendant matrix (\ref{r1234}) for $r=3$   would be defined on the product of the indecomposable representations $\mathcal{I}^8\otimes \mathcal{I}^8$.

 Consideration of the eigenproblem of Hamiltonian operators of the proposed integrable models with the help of the Quantum inverse scattering  method will require certain non-trivial extensions of the well developed methods of the Algebraic Bethe Ansatz (nested Bethe Ansatz), which is caused both by the reducible character of representations and by the complex structure of Lax operators.

  As it is known, by means of the YBE solutions the braid group representations can be realised, and they can be employed to obtain the link and knot invariants \cite{kr,TurJ,inv-satt}. Thus one  can use  the $R$-matrices defined on the composite spaces for determining link invariants for such extended cases too.  And besides of pure mathematical and theoretical interest, the solutions to Yang-Baxter equations on the reducible representations of the quantum algebras   also can have practical usage.  Particularly, such "extended" $R$-matrices %,
  can be attractive
    in the context of the recent developments of the mutual interrelations of the quantum entanglement theory (in topological aspects) and the theory of integrable models, or, more precisely, the solutions to YBE, as the essential instruments in the construction of integrable models \cite{LKL}. As the  Hamiltonian operators corresponding to the discussed solutions describe integrable systems having rather large number of degrees of freedom and rich structure of the spin variety (with quite tangled interactions) even for the lattices with few sites, so  possible applications may be assumed in different areas of 2d quantum statistical physics, string theories and particle physics.

\paragraph{Acknowledgement} The work is partly supported by the
Armenian Government Grant 15T-1C058.

\setcounter{section}{0}
\renewcommand{\thesection}{\Alph{section}}
\Alph{section}
\section{Appendix}
\renewcommand{\theequation}{\Alph{section}.\arabic{equation}}
\addtocounter{section}{0}\setcounter{equation}{0}
%\section{Appendix}
\subsection{ }
The action of the generators of the quantum super-algebra $osp_q(1|2)$ on the spin-$j_r$ irrep
$$V^r=\{v^r_{-\frac{r-1}{2}},v^r_{-\frac{r-3}{2}},\cdots
v^r_{\frac{r-1}{2}}\}$$ can be described by the following general
relations (see also \cite{shk}), $-\frac{r-1}{2}\leq
i\leq\frac{r-1}{2}$
\bea e\cdot v^r_{i}&=& \beta^r_i v^r_{i+1},\quad
\beta^r_{\frac{r-1}{2}}=0\nn,\\\label{efv} h \cdot
v^r_{i}&=&\left({i+[(-1)^r+1]\frac{i\pi}{4\log{q}}}\right)
v^r_{i},\\\nn f\cdot v^r_{i}&=&\gamma^r_{i} v^r_{i-1},\quad
\gamma^r_{-\frac{r-1}{2}}=0. \ena
The commutation relations of the algebra put the following
constraints on the coefficients $\alpha^r_i\equiv \beta^r_{i-1}
\gamma^r_{i}$, ($q_r=[(-1)^r+1]\frac{i\pi}{4\log{q}}$)
\bea \alpha^r_i=\sum_{i'=i}^{r-1}(-1)^{i'-i}[i'+q_r]_q=\frac
{(-1)^{r-1+i}[r/2+q_r]_q+[i+q_r-1/2]_q}{\sqrt{q}+1/\sqrt{q}},
\\
\alpha^{r}_{i}=-\alpha^{r}_{-i+1}.\ena
 The values of $\beta,\;
\gamma$ one can fix by normalizing the representation vectors.

\subsection{ }
Usually for finding the CG-coefficients the method
of the highest weight and the normalized vectors is used  (see for instance
\cite{kr}). Let us here  present
the coefficients in a general form \cite{shk}, with non-fixed
$\beta,\;\gamma$ coefficients in (\ref{efv}), for the particular
case $i=j$ %
\bea C\left(^{j_1 j_2\; j}_{i_1\; i_2\;
j}\right)=\prod_{i'_1=-j_1+1}^{i_1} \frac{-(-1)^{p_{
i'_1}}q^{-\frac{j+1+q_{r_1}+q_{r_2}}{2}}\beta^{r_1}_{i'_1-1}}{\beta^{r_2}_{j-i'_1}}
C\left(^{\;\;\;j_1\;\;\; j_2 \;\;\;j}_{-j_1\; j+j_1\; j}\right).
%\\
\ena
By convention we can suggest that the state with the highest weight has even grading.

Combining two relations, (\ref{v12}) and (\ref{12v}), and using
the orthogonality of the $v^r_i$-vectors, we obtain that $C$ and
$\bar{C}$-coefficients are inverse each to other in the following
matrix sense %(i.e. are forming inverse matrices to each other)
\bea
 \sum_{i_1=-j_1\{i_1+i_2=i\}}^{j_1}C\left(^{j_1\; j_2\; j}_{i_1\;
i_2\;\;i}\right)\bar{C}\left(^{j_1\; j_2\; j'}_{i_1\; i_2\;\;
i}\right)=\delta^{j j'}. \label{ccu}\ena
From the other hand, as the vectors $\{v^r_i\}$ and $\{v^{r'}_i\}$
are orthogonal when $r$ and $r'$ don't coincide, then it follows
that $\{C\left(^{j_1\; j_2\; j}_{i_1\; i_2\;\;
i}\right)\}\approx\{\bar{C}\left(^{j_1\; j_2\; j}_{i_1\; i_2\;\;
i}\right)\}$.
 The proportionality coefficients we can find from
the relation (\ref{ccu})
\bea\bar{C}\left(^{j_1\; j_2\; j\;}_{i_1\; i_2\;\;
i\;}\right)=(-1)^{p_{i_1}p_{i_2}}\varepsilon^j_i\varepsilon^{j_1}_{i_1}\varepsilon^{j_2}_{i_2}C\left(^{j_1\;
j_2\; j\;}_{i_1\; i_2\;\;i\;}\right),\label{cpr}\ena
where $\varepsilon^j_i$ is the norm of the state $v^r_i$. The norm
for the graded representations can be defined as in the work \cite{s}.
Let $v^r_j$ is an even state, then we can take $\varepsilon^j_i=1$
for all $i$. If $v^r_j$ is an odd state, then the norm in the
irrep $V_r$ is indefinite: $\varepsilon^j_i=(-1)^{j-i}$. For
definiteness we can take $v^{r_1}_{j_1}$ and $v^{r_2}_{j_2}$ as
even states, then the irreps $V^{r_1+r_2-1-k}$ have positive
norms, when $k=0+4\mathbb{Z}_{+}$ and have indefinite norms when
$k=0+2\mathbb{Z}_{+}$.

In the relations (\ref{efv}) the $\beta^r_{i-1},\;
\gamma^r_i$-coefficients  according to the mentioned
normalization can be fixed so, that
$\beta^r_{i-1}=\gamma^r_i(-1)^{j-i}$, and will be equal to
$\sqrt{\alpha^r_i}$ up to a sign.

\subsection{ }

\paragraph{Quantum algebra $sl_q(2)$.} At the end %
we give also brief definition of the quantum algebra $sl_q(2)$.
  Algebra generators are $e$,
$f$ and  $h$, %$k,\; k^{-1}$ ($k=q^h$),
 which satisfy to the following
commutation relations
 \bea
 [e,\;f]=\frac{q^h-q^{-h}}{q-q^{-1}},\quad   e\; q^{h+2}= q^h e,\quad f\;
q^h=q^{h+2} f.
 \ena
Co-product can be defined as
\bea \Delta[e]=e\otimes I+ q^h\otimes e,\quad \Delta[f]=f\otimes
q^{-h}+ I\otimes f,\quad \Delta[q^h]=q^h\otimes q^h.\ena
The quadratic Casimir operator is
\bea c=ef+\left(\frac{q^{h+1}-q^{-h-1}}{q-q^{-1}}\right)^2.\ena

Finite-dimensional irreducible representations $V^r$,
${\textrm{dim}}[V^r]=r$, are describing by their Casimir
eigenvalues $c_r=[r/2]_q^2$ and by "spin" values $j=(r-1)/2$,
with the analogy of the non-deformed algebra situation.

\paragraph{Quantum $6j$-symbols.} The associativity of the tensor product
of the quantum group  is expressed by definition of the quantum
$6j$-symbols $\left\{^{j_1 j_2 j_{12}}_{j_3 j\; j_{23}}\right\}_q$
as follows \cite{kr,grs}
\bea
\underline{{{_{j_1}}\;\;\;\Big{|}^{j_2}\;\;_{j_{12}}\;\;\Big{|}^{j_3}\;\;\;{_j}}}
=\sum_{j_{23}} \left\{^{j_1 j_2 j_{12}}_{j_3 j\;
{j_{23}}}\right\}_q
\underline{{\;{_{j_1}\;\;^{j_2}\overline{\;\;\;\;\;{\Big{|}\;_{j_{23}}}
\;}\;^{j_3}\;\;{_j}\;}}}\ena
Here $j,\;j_k$ are the spin values of the corresponding
$r$-dimensional irreps, $j=(r-1)/2$. The
first diagram corresponds to the tensor product $\{V_{j_1}\otimes
V_{j_2}{\;}_{\rightarrow}V_{j_{12}}\}\otimes V_{j_3}\rightarrow
V_j$, the second one in the sum corresponds to  $V_{j_1}\otimes
\{V_{j_2}\otimes V_{j_3}{\;}_{\rightarrow} V_{j_{23}}\}\rightarrow
V_j$.  Also let us use the notation $\rho(j_1,j_2;j)$ \cite{grs}
for denoting $c$-numbers which distinguish the Clebsh-Gordan
coefficients corresponding to the projections $V_{j}\to
V_{j_1}\otimes V_{j_2}$ and $V_j \to V_{j_2}\otimes V_{j_1}$. Then for revealing the spin structure of the operator (\ref{h}), one can
%there is
 consider as an elementary cell of the chain lattice (with sublattice structure) the product of the vector spaces
 $V_1\otimes V_2\otimes V_3 \otimes V_4$. And taking into account, that both of the terms of the Hamiltonian (\ref{h}, \ref{r1234}) contain the projector $P^1_{23}$,  we can compare the  following relations for the decomposition of the
vector products, in order to  express the terms containing the projectors
$P_{23}$ and $P_{14}$ acting on the product
$V_1\otimes\{V_2\otimes V_3\}\otimes V_4$, by means of the %some
 operators  acting
on the tensor product grouped as $\{V_1\otimes
V_2\}\otimes\{V_3\otimes V_4\}$
\bea&
\underline{{\;{_{j_1}\;\;^{j_2}\overline{\;\;\;\;\;{\Big{|}\;_{j_{23}}}
\;}\;^{j_3}\;\;{_{j_{123}}}\;}\;\;\;{\Big{|}}\;{}_{j_4}\;\;\;\;\;
{}_{j_{1234}}
}}=\sum_{{j_{12}},j_{34}}\frac{\rho(j_2,j_3,j_{23})}{\rho(j_1,j_{23},j_{123})}
\times&\nn\\&\rho(j_2,j_1,j_{12})\rho(j_3,j_{12},j_{123})
\left\{^{j_3\; j_2\;\;\; j_{23}}_{j_1\; j_{123}j_{12}}\right\}_q
\left\{^{j_{12}\;j_3\;\;j_{123}}_{j_4\;j_{1234}\;j_{34}}\right\}_q
\underline{{\;{_{j_1}}\;\;\;\Big{|}^{j_2}\;\;_{j_{12}}\;\;\;
^{j_3}\overline{\;\;\;\;\;{\Big{|}\;_{j_{34}}}
\;}\;^{j_4}\;\;\;{_{j_{1234}\;}}}}.&\ena
From the another hand we have
\bea\!\!\!
\underline{{\;{_{j_1}\;\;^{j_2}\overline{\;\;\;\;{\Big{|}\;_{j_{23}}}
}\;^{j_3}\;\;{_{j_{123}}}\;}\;\;\;{\Big{|}}\;{}_{j_4}\;\;\;\;
{}_{j_{1234}}
}}=\sum_{j_{14}}\rho(j_1,j_{23},j_{123})\{^{j_{23}\;j_1\;\;\;\;j_{123}}_{j_4\;\;j_{1234}\;j_{14}}\}_q\underline{{\;{_{j_2}}\;\;\;\Big{|}^{j_3}\;\;_{j_{23}}\;\;\;
^{j_1}\overline{\;\;\;\;{\Big{|}\;_{j_{14}}}
}\;^{j_4}\;\;\;{_{j_{1234}\;}}}}.\nn\\\hspace{-5cm}\ena
As the projector operator $P^1_{23}$ acting on the space
$V_{2}\otimes V_{3}$ maps it into the one dimensional space, then
$j_{23}=0$. And it means $j_{123}=j_1$, $j_{1234}=j_{14}$, so
$\{^{j_{23}\;j_1\;\;\;\;j_{123}}_{j_4\;\;j_{1234}\;j_{14}}\}_q\approx
\delta_{j_{14}, j_{1234}}$. These equations are valid for each variable
$j_{1234}$ which satisfies to $|j_{1}-j_{4}|\leq
j_{1234}\leq(j_1+j_4)$. The second term in the
$\check{R}_{1234}$-matrix contains the projector $P^{1}_{14}$,
%the action of
 which means %gives
 that for this case $j_{14}=0$, and hence $j_{1234}=0$ and
$j_{12}=j_{34}$. The %entire operators contain
external projectors
$({I}-P_{12}^1)({I}-P_{34}^1)$ entering into the
$\check{R}_{1234}$-matrix ensure that $j_{12}\neq 0$ and
$j_{34}\neq 0$. As all the states have the same dimension, i.e.
the same spin $j=(r-1)/2$, then $j_{12/34}\in [1,...,2j]$.

In the same spirit we can write out the transition operations passing the following steps:  $V_1\otimes V_2\otimes V_3\otimes V_4$ ${}_{\longrightarrow}^{P^E}$  $\sum\{V_1\otimes
V_2\}_{V_{12}}\otimes\{V_3\otimes V_4\}_{V_{34}}$ $\to$ $\sum V_1\otimes\{V_2\otimes V_3\}\otimes V_4$ ${}_{\longrightarrow}^{P^I}$
 $\sum \{V_1\otimes V_4\}_{V_{14}}\otimes\{V_2\otimes V_3\}_{V_{23}}$ ${}_{\longrightarrow}^{P^E}$ $\sum \{V_1\otimes
V_2\}_{V'_{12}}\otimes\{V_3\otimes V_4\}_{V'_{34}}$, which reflects the action of the Hamiltonian,
and here we denote the external and internal projection operators, entering into $H_{1234}$, as $P^E=(I-P^{1}_{12})(I-P^1_{34})$, $P^I=P^1_{23}$ or $P^I=P^1_{23}P^1_{14}$.
 In terms of the $6j$-symbols and the $\rho$-coefficients the action of the Hamiltonian term $H_{1234}$ would be obtained  from the following relation $\left(P^E P^I P^E\right)\times\;\sum_{j_{12},\;j_{34},\;j_{1234}}\underline{{\;{_{j_1}}\;\;\;\Big{|}^{j_2}\;\;_{j_{12}}\;\;\;
^{j_3}\overline{\;\;\;\;\;{\Big{|}\;_{j_{34}}}
\;}\;^{j_4}\;\;\;{_{j_{1234}\;}}}}$ $=$ $\sum_{j'_{12/34}\neq 0}\sum_{j_{12/34}\neq 0,\;j_{1234}\in[0,...,2j] }F_r(\rho)\prod\{\}_q\underline{{\;{_{j_1}}\;\;\;\Big{|}^{j_2}\;\;_{j'_{12}}\;\;\;
^{j_3}\overline{\;\;\;\;\;{\Big{|}\;_{j'_{34}}}
\;}\;^{j_4}\;\;\;{_{j_{1234}\;}}}}$, where $F_r(\rho)$ is a rational  function of the coefficients $\rho$, and the product of the quantum $6j$-symbols can be written as ($j_{1234}=j_{14}$) $\prod\{\}_q=\{^{j\;\; j\;\; j_{12}}_{j_{34}j_{14}j}\}_q\{^{j\; j\; j_{34}}_{j\;j\;\;0}\}_q\{^{0\; j\; j}_{j\;j_{14}j_{14}}\}_q\{^{j\; j\; j_{14}}_{0\;j_{14}\;j}\}_q\{^{j\; j\; 0}_{j\;j\;j'_{12}}\}_q\{^{j'_{12}\; j\;j}_{j\;j_{14}j'_{34}}\}_q$.

The explicit values of the quantum $6j$-symbols and numbers $\rho$
are calculated and can be %obtained
found  in the literature, see for example in \cite{grs}.

\end{document}